\documentclass[prc,twocolumn,showpacs,preprintnumbers,amsmath,amssymb]{revtex4-1}
\usepackage[dvips]{graphicx}
\usepackage{dcolumn}
\usepackage{bm}
\usepackage{color}

\begin{document}
\title{Status of Alpha-Particle Condensate Structure of the Hoyle State}

\author{Akihiro Tohsaki$^1$, Hisashi Horiuchi$^{1,2}$, Peter Schuck$^{3,4}$, Gerd R{\"o}pke$^5$}
\affiliation{$^1$Research Center for Nuclear Physics (RCNP), Osaka University, Ibaraki, Osaka 567-0047, Japan}
\affiliation{$^2$International Institute for Advanced Studies, Kizugawa, Kyoto 619-0225, Japan}
\affiliation{$^3$Institut de Physique Nucl$\acute{e}$aire, IN2P3-CNRS,
Universit$\acute{e}$ Paris-Sud, F-91406 Orsay Cedex, France}
\affiliation{$^4$Laboratoire de Physique et de Mod\'elisation des Milieux Condens\'es, CNRS, Grenoble Cedex 9, France, and
 Universit\'e Joseph Fourier, 25 Av. des Martyrs, BP 166, F-38042 }
\affiliation{$^5$Institut fuer Physik, Universitaet Rostock, Rostock 18051, Germany}

\begin{abstract}
The present understanding of the structure of the Hoyle state in $^{12}$C is reviewed. Most of the theoretical approaches to the Hoyle state are shortly  summarized. The corresponding results are analyzed with respect to whether they give evidence to the $\alpha$ particle condensation structure of the Hoyle state (and other Hoyle-like states in heavier self-conjugate nuclei) or not.
\end{abstract}
\maketitle
\section{Introduction}

The $0_2^+$ state at 7.65 MeV in $^{12}$C, known as the Hoyle state, is one of the most important states in nuclear physics. This stems from the fact that it is the gateway for the massive $^{12}$C production in the universe and is, thus, responsible for life on earth. It was predicted in 1954 by the the astrophysicist Fred Hoyle  \cite{Hoyle} at practically the correct energy (found by W. A. Fowler {\it et al.} in 1957 \cite{Fowler}). There was some discussion in the past whether F. Hoyle predicted his state on anthropic grounds or not. Apparently, this was not the case, see \cite{Kragh}. Standard shell model calculations give the energy of the first $0^+$ excited state in $^{12}$C at over 20 MeV. Therefore, because of its unexpected low energy, the structure of the Hoyle state stayed mysterious for a long time. In 1956  Morinaga postulated that it is a state where the three $\alpha$ particles are lined up straight, the so-called three $\alpha$ chain state \cite{Morinaga}. However, Horiuchi in 1974 \cite{Horiuchi} found with the semi-microscopic approach called Orthogonal Condition Model (OCM) \cite{Saito} (see below) that the Hoyle state should rather be interpreted as a state of three weakly coupled $\alpha$-particles interacting pair-wise in relative 0S-wave states. This point of view was confirmed in 1977 by two ground-breaking works by Kamimura {\it et al.} \cite{Kami} and Uegaki {\it et al.} \cite{Uegaki}. Using a phenomenological nucleon-nucleon force of Gaussian type which was adjusted independently earlier (Volkov force \cite{Volk}), they reproduced with a fully microscopic twelve nucleon wave function all known properties of the Hoyle and other loosely bound $\alpha$ states above the Hoyle state. The achievement of the two works was, at their time, so outstanding that- one is tempted to say that as usual after great exploits- the subject of the Hoyle state stayed practically dormant for about a quarter century. Only in 2001 appeared the work of Tohsaki, Horiuchi, Schuck, and R{\"o}pke (THSR) which interpreted the Hoyle state (and other states in self-conjugate nuclei) as a condensate of $\alpha$-particles \cite{thsr}. That means that the $\alpha$ particles with their center of mass motion occupy all the lowest 0S orbit of their common mean field. This work triggered an intense new interest in the Hoyle state, both theoretically and experimentally, see the review articles in \cite{reviews} and references in there. At present, this research culminates in works trying to explain the properties of the Hoyle state from {\it ab initio} and/or Quantum Monte Carlo (QMC) approaches \cite{Wiringa, Meissner}. Already in the work of THSR \cite{thsr}, it was predicted that in other $n\alpha$ nuclei Hoyle-analog states should exist around the $n\alpha$ disintegration threshold. For instance $^{16}$O is subject of intense studies, both theoretically and experimentally \cite{Itoh, Itoh2, OCM16, Meissner2}. The situation in these self-conjugate nuclei is now considered under a completely novel aspect, namely that at energies, close to the $\alpha$ disintegration threshold, there exist states of extended volume (3-4 times the volume of the ground state) where the nuclei are formed by a gas of loosely bound $\alpha$ particles which move in their own mean field. These bosonic states co-exist with the standard fermionic ones, where individual nucleons move in a common mean field. This is a very exciting new feature of nuclear physics, of importance for nuclei and for astrophysical aspects.

Pairing is well known and well accepted as a very useful approximation to describe two particle correlations. 
Pairing between like nucleons (n-n, p-p) is a useful concept 
not only
in infinite matter, but also in finite nuclei \cite{RS}.
Because the interaction between protons and neutrons (deuteron channel)
is even stronger (a bound state can be formed), one should expect that 
in symmetric matter also proton-neutron pairing should appear.
Calculations performed for the p - n channel give a critical temperature
depending on density which can rise up to a value $T_c$ near to 4.5 MeV
\cite{Meng} which is about three times 
larger than the maximum critical temperature in the isospin triplet
channel.
However, it competes with the formation of $\alpha$ particles,
and for these particles the transition to a Bose condensate at increasing
density
occurs prior to the quantum condensation in the deuteron channel  as
discussed below.

The next step is quartetting which is also a very good approximation in special situations. The THSR wave function was an important step to be introduced in nuclear structure physics.
Nuclear physics is now on the forefront of the studies on quartet and cluster formation, and their possible condensation, but there are also works in other fields 
with still ongoing interest presently. For example there exist speculations that in semiconductors where excitons, that is bound states of a conduction electron and an electron hole, can bind to bi-excitons which may enter in competition with single excitons in a possible Bose-Einstein condensation (BEC) \cite{Noz}. There exist also theoretical works which predict that, once four different fermions can be trapped in cold atom devices, bound quartets can be formed with again the possibility of BEC \cite{Lechem}.
 
The purpose of this paper is to  review briefly the present situation concerning the possibility of $\alpha$ particle condensation and  other $\alpha$ gas states in selfconjugate nuclei which was proposed for the first time 15 years back in \cite{thsr}. It may be important to clarify already at this point that we  understand the word ``condensate'' or ``condensation'' in the sense that the $\alpha$ particles with inert internal structure move all with their center of mass (c.o.m.) in the same lowest 0S orbit of their common mean field. We will show in the main part of this article that these states can be considered as the precursors of a macroscopic $\alpha$ particle condensate in homogeneous nuclear matter at low density. Of course, the reader should understand that in the following the term 'condensate' for a handful of $\alpha$ particles is stretched to its limits.\\ 
Because of its outstanding importance, we will mainly concentrate on $^{12}$C and the Hoyle state but, at the end of this colloquium, we also will touch the situation in other nuclei. We shall present in condensed form the various theoretical attempts which are used to describe the formation and existence of quartets in nuclei, that is the $\alpha$ gas states, induced by strong four nucleon correlations (2 neutrons-2 protons) at densities well below saturation. We will discuss to which degree they give arguments against or in favor of the hypothesis that the Hoyle state can be considered to good approximation as a state where the three $\alpha$'s are condensed into the c.o.m. 0S-orbital. \\

Historically, the idea of $\alpha$ condensation is based on the study in ref. \cite{Roepke} where the critical temperature of quartet condensation in infinite matter was investigated. This study was performed in analogy to the determination of the critical temperature for the onset of pairing, i.e., superfluidity or superconductivity, employing the in-medium two fermion equation as done by Thouless \cite{Thouless} (now known as the Thouless criterion). For the $\alpha$ particle, the corresponding in-medium four fermion equation has been used and solved in \cite{Roepke}. Since the $\alpha$ particle is a very strongly bound quartet with a binding energy/particle  of $\sim$ 7.5 MeV which is about seven times  larger than the one of the deuteron and almost as large as in the strongest bound nucleus which is Iron, the critical temperature turned out to be, at low density, over a factor of six higher than the one of neutron-neutron pairing.
This finding was then logically transposed, in analogy what had happened in the case of pairing, to finite nuclei. The presentation of the physics involved in quartetting is the main subject of this colloquium.

The paper is organized as follows. In Sect.II, we give a short summary of the THSR approach with the main focus on what it predicts with respect to the Hoyle state being an $\alpha$-particle condensate state. In Sect. III, we revisit in a nutshell all other theories which may have some connection with the $\alpha$ condensate aspect. In Sect. IV we give a glimpse on the situation in $^{16}$O and in Sect. V, eventual experimental evidences are discussed. Finally in the last section, we present some further discussions together with our conclusions and a short outlook.

\vspace{1.5cm}
\section{ The THSR approach and the Hoyle state}

\vspace{0.5cm}
As already mentioned, in 2001 a new aspect of the Hoyle state came at the forefront of the discussion. In \cite{thsr} it was advanced that the Hoyle state might not only be a gas-like state of three $\alpha$-particles but it was suggested that the three $\alpha$'s are, with their center of mass (c.o.m.) motion condensed into an identical 0S-orbital, a situation reminiscent of what happens in cold atom physics where, however, exists a much larger number of bosons. In addition it was predicted that not only $^{12}$C contains such an $\alpha$ condensate but also several heavier self-conjugate nuclei like $^{16}$O, $^{20}$Ne, etc. may exhibit analogous features. The idea of condensation was first investigated in nuclear matter \cite{Roepke} and then borne out by the use of a condensate type of wave function for finite self-conjugate nuclei, now known according to their authors as the THSR wave function \cite{thsr}. The most basic form of THSR is, e.g., for the case of three $\alpha$ particles a {\it single} wave function of following structure

\begin{equation}
\Psi_{\mbox{THSR}} \propto {\mathcal A} \psi_1\psi_2\psi_3 \equiv {\mathcal A}|B\rangle
\label{THSRwf}
\end{equation}

\noindent
with

\begin{equation}
\psi_i = e^{-(({\bf R}_i-{\bf X}_G)^2)/B^2}\phi_{\alpha_i}
\label{a-wf}
\end{equation}

\noindent
and

\begin{equation}
\phi_{\alpha_i} = e^{-\sum_{k<l}({\bf r}_{i,k}-{\bf r}_{i,l})^2/(8b^2)}
\label{int-a-wf}
\end{equation}

\noindent
In (\ref{THSRwf}) the ${\bf R}_i$ are the c.o.m. coordinates of $\alpha$ particle '$i$' and ${\bf X}_G$ is the total c.o.m. coordinate of $^{12}$C. ${\mathcal A}$ is the antisymmetrizer of the twelve nucleon wave function with $\phi_{\alpha_i}$ the intrinsic translational invariant wave function of the $\alpha$-particle '$i$'. The whole 12 nucleon wave function in (\ref{THSRwf}) is, therefore, translationally invariant. Please note that we suppressed the scalar spin-isospin part of the wave function. The special Gaussian form given in Eqs.  (\ref{a-wf}), ( \ref{int-a-wf}) was chosen in \cite{thsr} to ease the variational calculation. The condensate aspect lies in the fact that (\ref{THSRwf}) is a (antisymmetrized) product of three times the same $\alpha$-particle wave function and is, thus, analogous to a number projected BCS wave function in the case of pairing.  This twelve nucleon wave function has 
two variational parameters, $b$ and $B$. It possesses the remarkable property that for $B=b$ it is a pure harmonic oscillator Slater determinant (this aspect of (\ref{THSRwf}) is explained in \cite{B-Bohr, Yam08}) whereas for $B \gg b$ the $\alpha$'s are at low density so far apart from one another that the antisymmetrizer can be dropped and, thus, (\ref{THSRwf}) becomes a simple product of three $\alpha$ particles, all in identical 0S states, that is, a pure condensate state. The minimization of the energy with a Hamiltonian containing a nucleon-nucleon force determined earlier independently \cite{Toh-force} allows to obtain a reasonable value for the ground state energy of $ ^{12}$C. Variation of energy under the condition that (\ref{THSRwf}) is orthogonal to the previously determined ground state allows to calculate the first excited $0^+$ state, i.e., the Hoyle state. While the size of the individual $\alpha$ particles remains very close to their free space value ($b \simeq$ 1.37 fm), the variationally determined $B$ parameter takes on about three times this value. It is important to mention right away that this so determined THSR wave function has about 98 percent squared overlap with the one of Kamimura {\it et al.} \cite{Kami} (and practically 100 percent squared overlap with a slightly more general THSR wave function superposing several $B$ values). We will shortly explain Kamimura's wave function below in Sect. III. It can, even to day, after 40 years, be considered as one of the most efficient approaches for the Hoyle state. In any case, as in the work of Kamimura {\it et al.} \cite{Kami}, so does the THSR approach reproduce very well all known experimental data about the Hoyle state. This concerns for instance the inelastic form factor, electromagnetic transition probability, and position of energy, see for more details \cite{reviews} and \cite{inelastic}. The inelastic form factor is shown in  Fig.\ref{QMC} below. At practically the same time Uegaki {\it et al.} published a very similar paper leading to almost identical results \cite{Uegaki}. In the following, we often will only refer to Kamimura's work, since we were able to compare THSR and Kamimura's wave functions numerically. However, all what we say below about Kamimura's work should equally apply to the one of Uegaki.

The THSR wave function contains two limits: a pure Slater determinant and a pure Bose condensate below about a 5-th of the saturation density $\rho_0$ . To which end, the Hoyle state is closest? 

\begin{figure}[h]
\begin{center}
\includegraphics[scale=0.65]{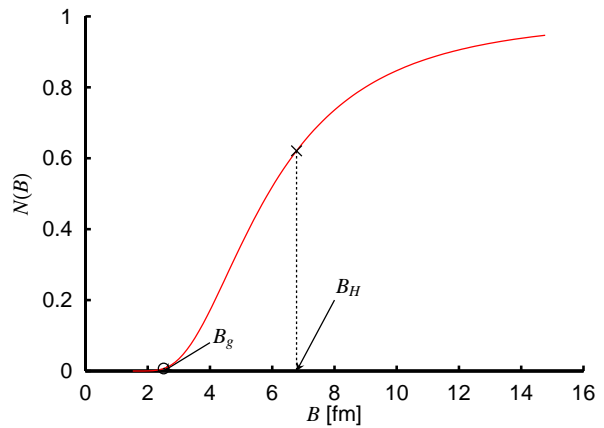}
\caption{Expectation value of the antisymmetrization operator in the product state $|B\rangle$. The value at the optimal $B$ values, $B_g$ for the ground state and $B_H$ for the Hoyle state, are denoted by a circle and a cross, respectively.}
\label{fig:anti}
\end{center}
\end{figure}


To this end it is very instructive to consider the effect of the antisymmetriser in (\ref{THSRwf}) in more detail. In Fig.\ref{fig:anti} we show the expectation value of the antisymmetrizer 

\begin{equation}
N(B) = \frac{ \langle B|{\mathcal A}|B\rangle}{\langle B|B\rangle}
\label{anti-sym}
\end{equation}
in the Hoyle state. Indicated are the optimal values of the $B$ parameter for the ground state and the Hoyle state.

\begin{figure}\begin{center}
\includegraphics[height=6cm]{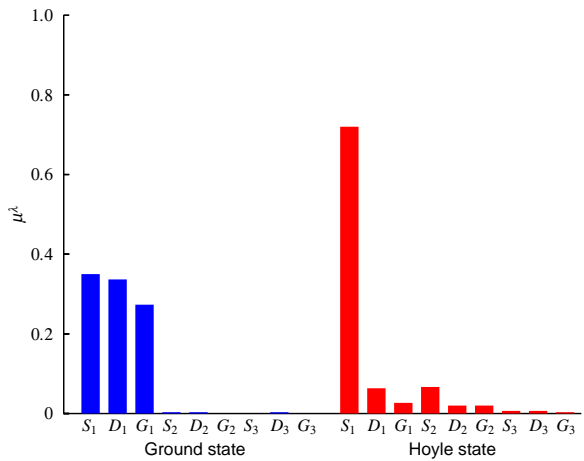}
\end{center}
\caption{$\alpha$ particle occupation numbers in the ground state (left) and in the Hoyle state (right) \cite{occ's}}.
\label{occs-12C}
\end{figure}

For $B \rightarrow \infty$ the quantity in (\ref{anti-sym}) tends to one, since, as already mentioned, the $\alpha$ particles are in this case so far apart from one another that antisymmetrisation becomes negligeable. The result for $N(B)$ is shown in Fig.~\ref{fig:anti} as a function of the width parameter $B$. We chose as optimal values of $B$ for describing the ground and Hoyle states, $B=B_g = 2.5$ fm and $B=B_H$= 6.8 fm, for which the normalised THSR wave functions give the best approximation of the ground state $0^+_1$ and the Hoyle state $0^+_2$, respectively. We find that $N(B_H) \sim 0.62$ and $N(B_g) \sim 0.007$. These results indicate that the influence of the antisymmetrisation is strongly reduced in the Hoyle state compared with the ground state. This study gives us a first indication that the Hoyle state is quite close to the quartet condensation situation rather than being close to a Slater determinant. 
However, there is another quantity which tells us more directly whether the Hoyle state is close to a three $\alpha$ condensate or not.
 In \cite{Suzuki} Suzuki {\it et al.} evaluated the bosonic occupation numbers using a Gaussian representation of the c.o.m. part $\chi$ of the RGM (Resonating Group Method) wave function, see Sect. III.B,  to calculate the
 single $\alpha$ particle density matrix $\rho_{\alpha}({\bf R},{\bf R}')$ and diagonalising it. The bosonic occupation numbers were also calculated by Yamada et al. in \cite{occ's} using, however, the already mention OCM approach, see below. Both calculations concluded that the three $\alpha$'s in the Hoyle state occupy to about 70 percent the same 0S orbit whereas all other ones are down by at least a factor of ten, see Fig.~\ref{occs-12C}. The density-induced suppression of the $\alpha$ particle condensate has also been studied in \cite{Gerd}.

The THSR calculation also showed that the inelastic form factor from the ground to Hoyle state is very sensitive to the size of the Hoyle state ~\cite{inelastic}. For example extending artificially the size of the Hoyle state by 20$\%$, increases the inelastic form factor globally by a factor of two. The THSR approach yields for the rms radius 3.83 fm (ground state 2.4 fm), so that the volume (density) of the Hoyle state is approximately 3-4 times larger (lower) than the one of the ground state. Those numbers are rather similar to what one finds for $^8$Be reinforcing the picture of the Hoyle state of a low density three $\alpha$ system where the $\alpha$ particles are individually well borne out, see Fig.~\ref{densities8Be} below. Since the THSR wave function has 98-100 $\%$ squared overlap (depending on more or less elaborate versions of THSR) with the wave function for the Hoyle state of Kamimura {\it et al.} \cite{Kami} which, together with \cite{Uegaki} can be considered as the most general ansatz used so far with practically perfect precision even far out in the tail, one can deduce that implicitly the Kamimura approach also gives a $\sim$ 70$\%$ bosonic occupancy for the $\alpha$ particles  in the Hoyle state. As a remark on the side, one may notice that for single proton or neutron states in nuclei one also obtains occupancies of 70-80$\%$ ~\cite{Frois}. One may, therefore, say that the bosonic quartets in nuclei excited to energies around the $\alpha$ decay threshold are about as far from (or as close to) the ideal gas case as are the fermions in the ground state. On the other hand, in cold atom devices, the bosonic atoms are at so low densities that their electronic clouds do not overlap at all and, thus, ideal Bose-Einstein condensation (BEC) can develop ~\cite{String}.\\ 

At this point, let us stress again that terms like '$\alpha$ particle condensation' or 'Bose-Einstein condensation' strictu sensu apply only for macroscopic systems as homogeneous nuclear matter which we will treat later. In finite nuclei such terms can only be used in the sense that 'condensate states' are to be considered as precursers to what happens potentially in the infinite matter case.

\section{Further approaches to the Hoyle state}

\subsection{OCM of Horiuchi {\it et al.}}

The precursor of all more or less realistic tentatives to explain the Hoyle state is the semi-microscopic description by Horiuchi {\it et al.} in 1975 \cite{Horiuchi} using the OCM approach as mentioned in the Introduction. In the latter, the $\alpha$ particles are replaced by ideal bosons interacting with phenomenological two and three body forces. However, in an important aspect the Pauli principle is incorporated into the OCM approach. It is related to the fact that the physical states should be orthogonal to the so-called Pauli forbidden states. So, in OCM the 2, 3, ... body bosonic equations are solved under the condition $\langle u_{\mbox{F}}|\Phi_{\mbox{OCM}}\rangle = 0$ where $u_{\mbox{F}}(r)$ are the Pauli forbidden
states. For example in the case of $^8$Be those are given by harmonic oscillator 0S, 1S, and 0D wave functions (up to four $\hbar \omega$ quanta) because four neutrons plus four protons in a harmonic oscillator also occupy four $\hbar \omega$ quanta. In \cite{Horiuchi}, it is stated for the first time that the Hoyle state is not a linear chain state but rather a state of {\it '3$\alpha$'s weakly coupled to each other in relative S-states'}. It also was concluded that the Hoyle state has quite enlarged spacial structure compared to the one of the ground state of $^{12}$C. The authors did not investigate the Bose condensate character of the Hoyle state but it is clear that from a state of {\it 'loosely bound $\alpha$ particles'} to a condensate of $\alpha$'s, there is only a short step.

\begin{figure}
\includegraphics[width=8cm]{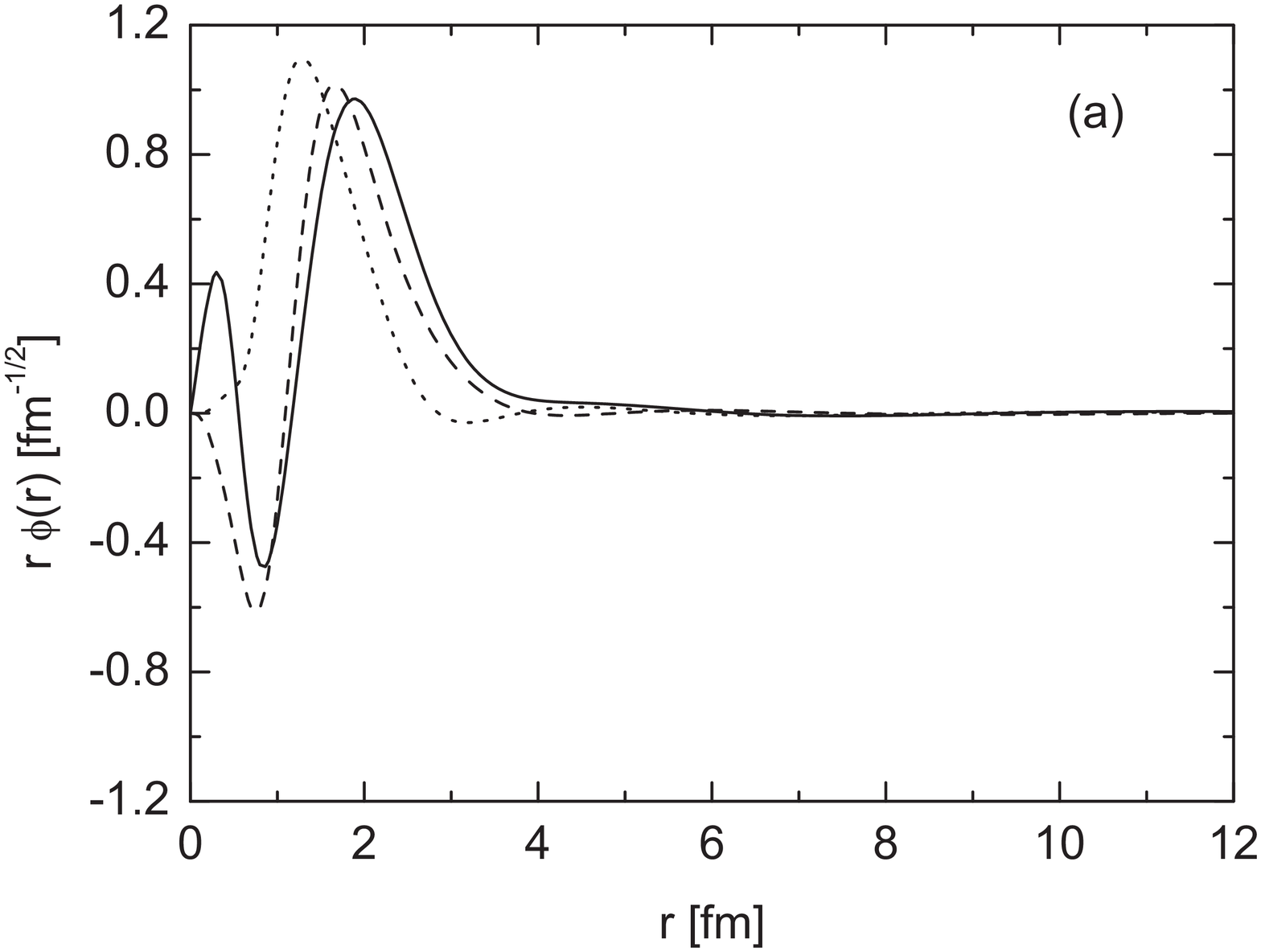}
\includegraphics[width=8cm]{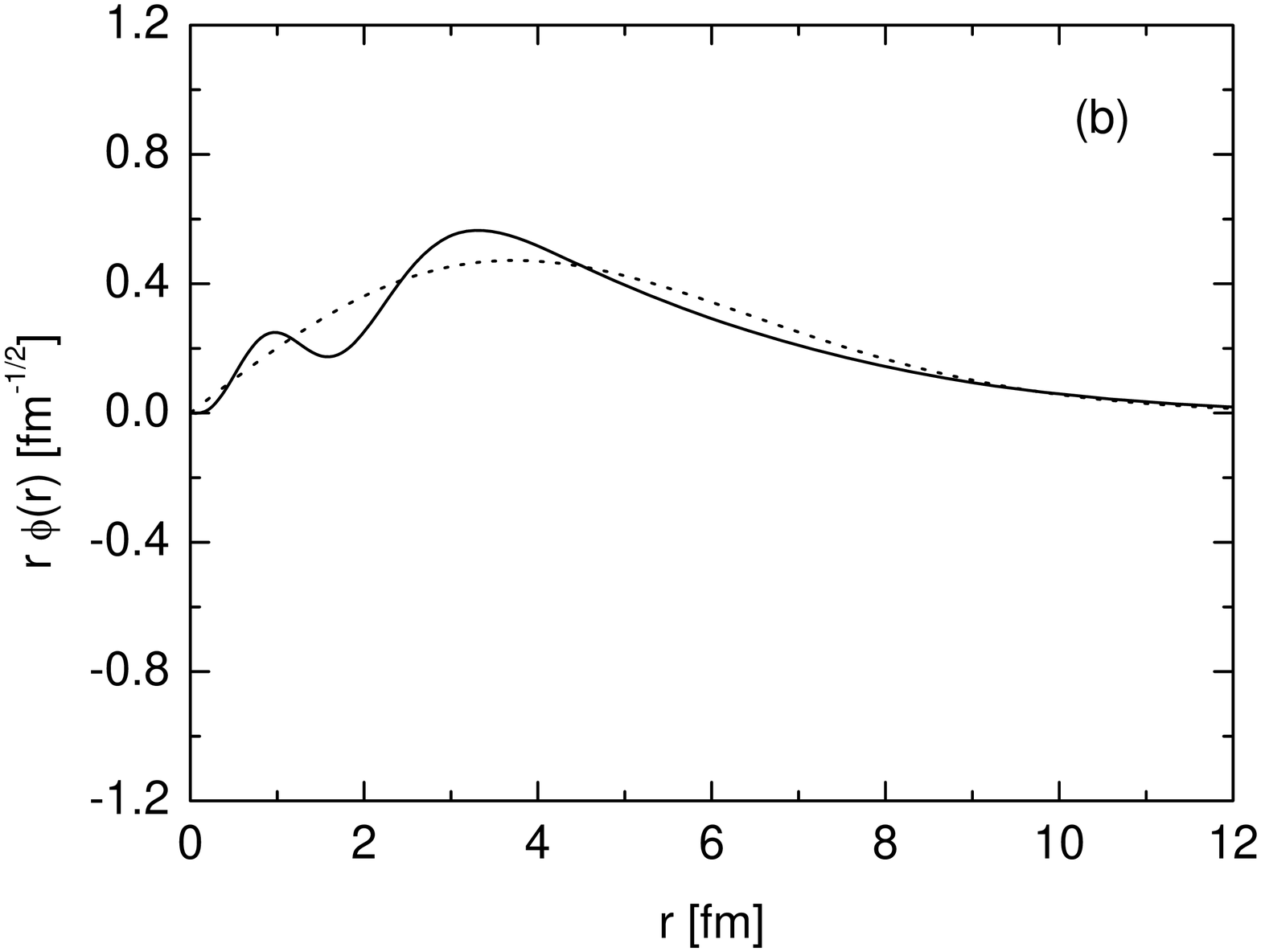}
\caption{\label{gs-wf12C}
The ground state wave function (upper panel) is compared to the one of the Hoyle state (lower panel). We see the strong difference in spacial extensions. The strong overlap of $\alpha$'s in the ground state is responsible for the pronounced oscillations (upper panel) whereas in the Hoyle state the S-wave function resembles a broad Gaussian. In the upper panel, the full line corresponds to the S-wave, broken line D-wave, and dotted line G-wave.}
\end{figure}

As we will see later in Sect. IV, concerning a study of the $0^+$ spectrum in $^{16}$O, OCM remains a very efficient method for $\alpha$ cluster states. We show in Fig.~\ref{gs-wf12C} the radial part of the Hoyle wave function calculated with OCM \cite{occ's} (full line). We see no nodal behavior of the Hoyle orbit, only small oscillations in the inner region and a long tail up to $r \sim$ 10 fm. The radial behavior of the Hoyle orbit is similar to a Gaussian  (dotted line). On the contrary, in the ground state of $^{12}$C where the $\alpha$'s strongly overlap, due to the active Pauli principle, strong oscillations develop with number of nodes two, one, and zero for S, D, and G waves, respectively. This reflects very well the SU(3) character of the $^{12}$C ground state.

\subsection{The approaches by Kamimura {\it et al.} and by Uegaki {\it et al.}}

Kamimura {\it et al.} \cite{Kami} made the following RGM ansatz for the Hoyle state

\begin{equation}
\Psi_{\mbox{RGM}} \propto {\mathcal A}\chi({\bf \xi}_1,{\bf \xi}_2)\phi_{\alpha_1}\phi_{\alpha_2}\phi_{\alpha_3}
\label{Kamimura}
\end{equation}

\noindent
whereas Uegaki et al \cite{Uegaki} considered the  Brink-GCM (Generator Coordinate Method) wave function

\begin{eqnarray}
\Psi_{\mbox{B-GCM}} &\propto& P_0\int d^3{\bf S}_1\int d^3{\bf S}_2\int d^3{\bf S}_3f({\bf S}_1,{\bf S}_2,{\bf S}_3)\Phi_{\mbox{B}}\nonumber\\
\Phi_{\mbox{B}}=&{\mathcal A}&e^{-({\bf R}_1 - {\bf S}_1)^2/b^2}e^{-({\bf R}_2 - {\bf S}_2)^2/b^2}e^{-({\bf R}_3 - {\bf S}_3)^2/b^2}\nonumber\\
&&\phi_{\alpha_1}\phi_{\alpha_2}\phi_{\alpha_3}
\label{Uegaki}
\end{eqnarray}

\noindent
where $\phi_{\alpha_i}$ is again the intrinsic $\alpha$ particle wave function of (\ref{int-a-wf}), ${\bf \xi}_i$ are Jacobi coordinates and $\chi({\bf \xi}_1,{\bf \xi}_2)$ is a completely general translational invariant three boson wave function (please note again that in Eqs.~(\ref{Kamimura}), (\ref{Uegaki}) the scalar spin-isospin part of the wave function is not written out). In (\ref{Uegaki}) $ f({\bf S}_1,{\bf S}_2,{\bf S}_3)$ is  the generator coordinate weight function and $P_0$ is a projector onto zero total momentum. The Brink wave function $\Phi_{\mbox{B}}$ \cite{Brink} places the small sized $\alpha$ particles with width parameter $b$ at definite spatial points ${\bf S}_i$. These three body
wave functions have been determined variationally with RGM (GCM) by Kamimura et al.\cite{Kami} (Uegaki et al.\cite{Uegaki}). Notice that the angle of the third $\alpha$ with respect to the axis of the other two is completely free, as well as the distance with respect to the other two $\alpha$ particles. Therefore, all kinds of 3$\alpha$ arrangements from linear chain over open triangle to equilateral triangle, etc. are in principle possible. On the other hand, since the THSR wave function is equivalent to  Kamimura's wave function, this tells us that implicitly Kamimura's wave function also contains to about 70 $\%$ an $\alpha$ particle condensate component. We will later show in Fig.\ref{QMC} that the inelastic form factor of Kamimura and THSR are on top of one another  explaining very accurately the experimental data. In addition to all known properties of the Hoyle state, Kamimura et al. and Uegaki et al. explained a variety of other $\alpha$ gas states in $^{12}$C with different quantum numbers such as, e.g., the second 2$^+$ state whose position was experimentally only confirmed very recently \cite{Itoh3}. As Horiuchi, also Kamimura and Uegaki concluded that the Hoyle state is a {\it 'weakly coupled system or gas of alpha particles'}   in relative S-waves. We cite Uegaki {\it et al.} \cite{Uegaki}: {\it ' In a number of excited states which belong to the new ``phase'', $^{12}$C nucleus should be considered to dissociate into 3 $\alpha$-clusters which interact weakly with each other and move almost freely over a wide region.'} And further: {\it 'The $0_2^+$ state is the lowest state which belongs to the new ``phase'', and could be considered to be a finite system of $\alpha$-boson gas.'} 
These words are very similar to what we use nowadays in the context of the THSR approach. For instance, the 'container' picture \cite{Bo2} of which we again will talk later in Sect.III.E, is already alluded to. The major difference between THSR and those earlier works consists in that THSR predicted that $\alpha$ particle condensation may not only exist in $^{12}$C but also in heavier $n\alpha$ nuclei, as, for instance in $^{16}$O \cite{thsr} and, thus, may be a more  general phenomenon. Also the bosonic occupation numbers were not calculated at that time.

\subsection{AMD and FMD approaches by Kanada En'yo {\it et al.} and Chernykh {\it et al.}}

In 2007 the Hoyle state was also newly calculated by the practioneers of Antisymmetrized Molecular Dynamics (AMD) (Kanada En'yo {\it et al.} \cite{Ishiko}) and Fermion Molecular Dynamics (FMD) (Chernykh {\it et al.} \cite{Richter}) approaches. In AMD one uses a Slater determinant of single-particle Gaussian wave packets where the center of the packets ${\bf S}_i$ are replaced by complex numbers. This allows to give the center of the Gaussians a velocity as one easily realizes. In FMD in addition the width parameters of the Gaussians are also complex numbers and, in principle, different for each nucleon. AMD and FMD do not contain any preconceived information of clustering. Both approaches found from a variational determination of the parameters of the wave function and a prior projection on good total linear and angular momenta that the Hoyle state has dominantly a 3-$\alpha$ cluster structure with no definite geometrical configurations. In this way the $\alpha$ cluster ans\"atze of the earlier approaches were justified. 
As an  achievment, in \cite{Richter}, the inelastic form factor from the ground to Hoyle state was successfully reproduced in employing an effective nucleon-nucleon interaction V$_{\mbox{UCOM}}$ derived from the realistic bare Argonne V18 potential (plus a small phenomenological correction).


Kanada En'yo {\it et al.} \cite{Ishiko} pointed out that with AMD some breaking of the $\alpha$ clusters can and is taken into account. The Volkov  force \cite{Volk} was employed in \cite{Ishiko}. Again all properties of the Hoyle state were explained with these approaches. Like in the other works \cite{Kami, Uegaki, Richter, thsr}, the E0 transition probability came out $\sim$ 20 $\%$ too high. No bosonic occupation numbers were calculated. It seems technically difficult to do this with these types of wave functions. However, one can suspect that if occupation numbers were calculated, the results would not be very different from the THSR results. This stems from the high sensitivity of the inelastic form factor to the employed wave function.  Nontheless, it would be important to produce the occupation numbers also with AMD and FMD.

In \cite{Ishiko, Richter} some geometrical configurations of $\alpha$ particles in the Hoyle state are shown. No special configuration out of several is dominant. This reflects the fact that the Hoyle state is not in a crystal-like $\alpha$ configuration but rather forms to a large extent a Bose condensate. 

\subsection{Pure bosonic approaches}

In some works the Hoyle state is approached in treating the $\alpha$ particles completely as ideal bosons. Even the fact that the physical states should be orthogonal to the Pauli forbidden states, as is done in OCM, is not taken care of. The effect of antisymmetrisation is entirely simulated by effective forces. 
The two most recent approaches of this sort are the ones of Lazauskas {\it et al.} \cite{Rimas} (who used the non-local Papp-Moszkowski force \cite{Papp}) and, more recently, of  Ishikawa \cite{Ishikawa} using a modified Ali-Bodmer interaction ~\cite{AliB} plus a three body term. In \cite{Rimas, Ishikawa} the position of the Hoyle state and the $\alpha$ threshold energies are well reproduced. In \cite{Rimas} also the relative angular momenta  between the $\alpha$ particles in the Hoyle states are analysed. It turns out that there is S-wave dominance to about 80 $\%$. This implies that also the S-wave occupation number is of the order of 70-80 $\%$. This is shown in \cite{Ishikawa} where such an analysis was also performed. It was found that the proportion of partial waves is practically the same as in \cite{Rimas}. In addition the bosonic occupation numbers were calculated and the 0S occupancy turned out to be $\sim$ 80 $\%$, thus confirming the Bose-condensate picture \cite{Ishi-private}. The strong link between relative S-wave dominance and high S-wave occupation numbers is likely a general feature. On the other hand in \cite{Ishikawa} the simultaneous and democratic three $\alpha$ decay probability was given. This can be considered as a great achievement. The probability with respect to sequential $^8$Be + $\alpha$ decay resulted to be negligeable (branching ratio: 10$^{-4}$). However, this does not speak against $\alpha$ particle condensation. It simply means that three body decay (tunnelling under the Coulomb barrier) is strongly hindered.

\subsection{Brink-type versus THSR wave function. Dumbbell {\it vs.} container picture}

A nice way to compare Brink and THSR wave functions is the following hybrid ansatz, e.g., for $^8$Be

\begin{eqnarray}
\Psi_{\mbox{THSR-hyb}} &\propto &P_0{\mathcal A}e^{-({\bf R}_1 - {\bf S}_1)^2/B^2}e^{-({\bf R}_2 - {\bf S}_2)^2/B^2}\nonumber\\
&&\times \phi_{\alpha_1}\phi_{\alpha_2}
\label{hybrid}
\end{eqnarray}

\noindent
In this way the THSR and Brink wave functions are encapsulated in one formula. For $B=b$, we have the Brink wave function $\Phi_{\mbox{B}}$ (\ref{Uegaki}) and for $S_i=0$, we have the THSR wave function (\ref{THSRwf}). It turned out in a number of examples where the two variational parameters $B$ and $S$  have been put into competition that always $B > b$ and $S=0$ was the outcome of a variational calculation, see, e.g., \cite{Bo} and unpublished work. Therefore, the THSR picture where the large $B$ parameter indicates free mean field motion of the cluster, the so-called 'container picture' \cite{Bo2} prevails over the Brink ansatz where the clusters are nailed down to definite positions via the ${\bf S}_i$ parameters. This latter evokes the 'dumbbell' or 'molecular' picture which was used almost exclusively in the past, for example in the description of $^8$Be. It is true that most of the time, not a single Brink wave function was considered but a superposition smearing out the position of the clusters. This was believed to be a correction and the underlying picture was thought to remain the dumbbell or molecular one. However, as the studies with the hybrid wave function (\ref{hybrid}) show, the basic property of cluster motion is just the contrary: free motion in a cluster mean field, the 'container'. Of course, the clusters in their motion cannot penetrate each other, due to the Pauli principle. 
\begin{figure}
\includegraphics[width=6cm,angle=-90]{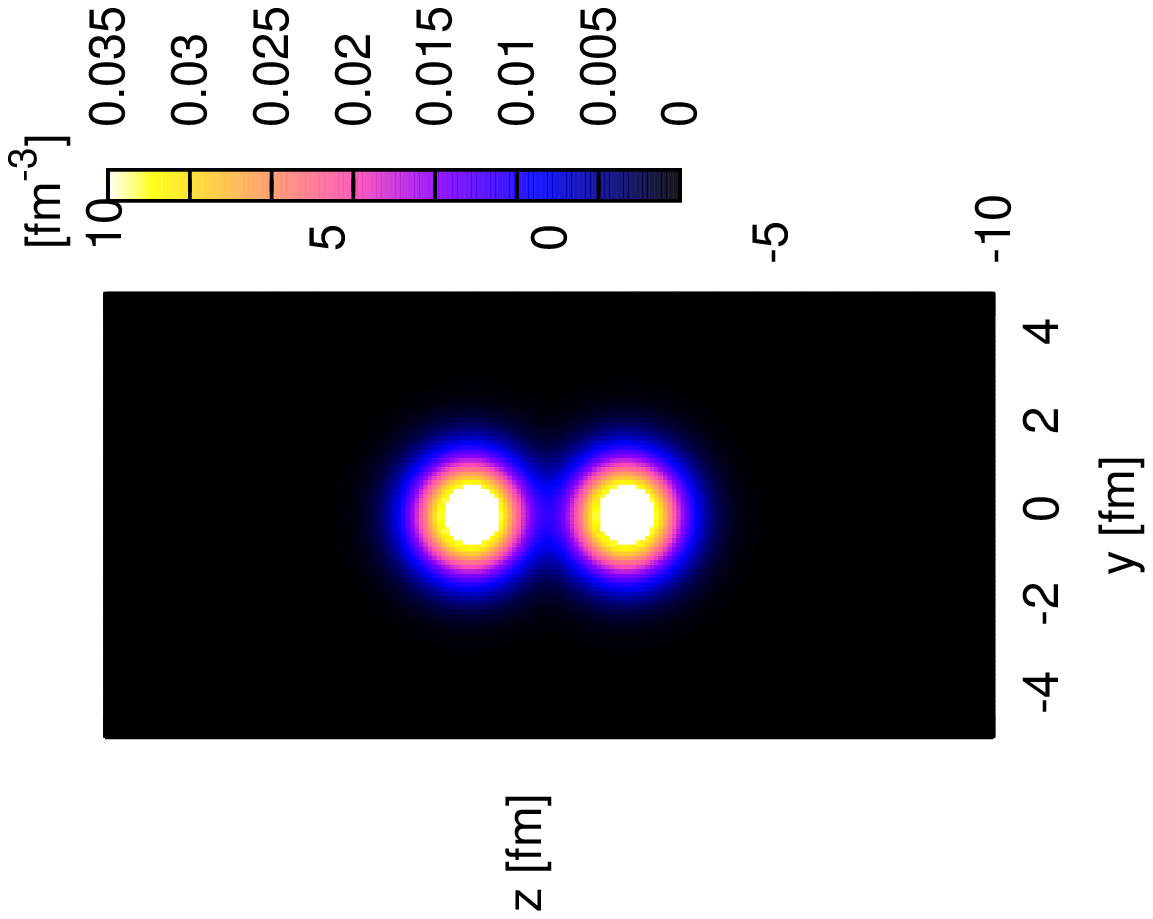}
\includegraphics[width=6cm,angle=-90]{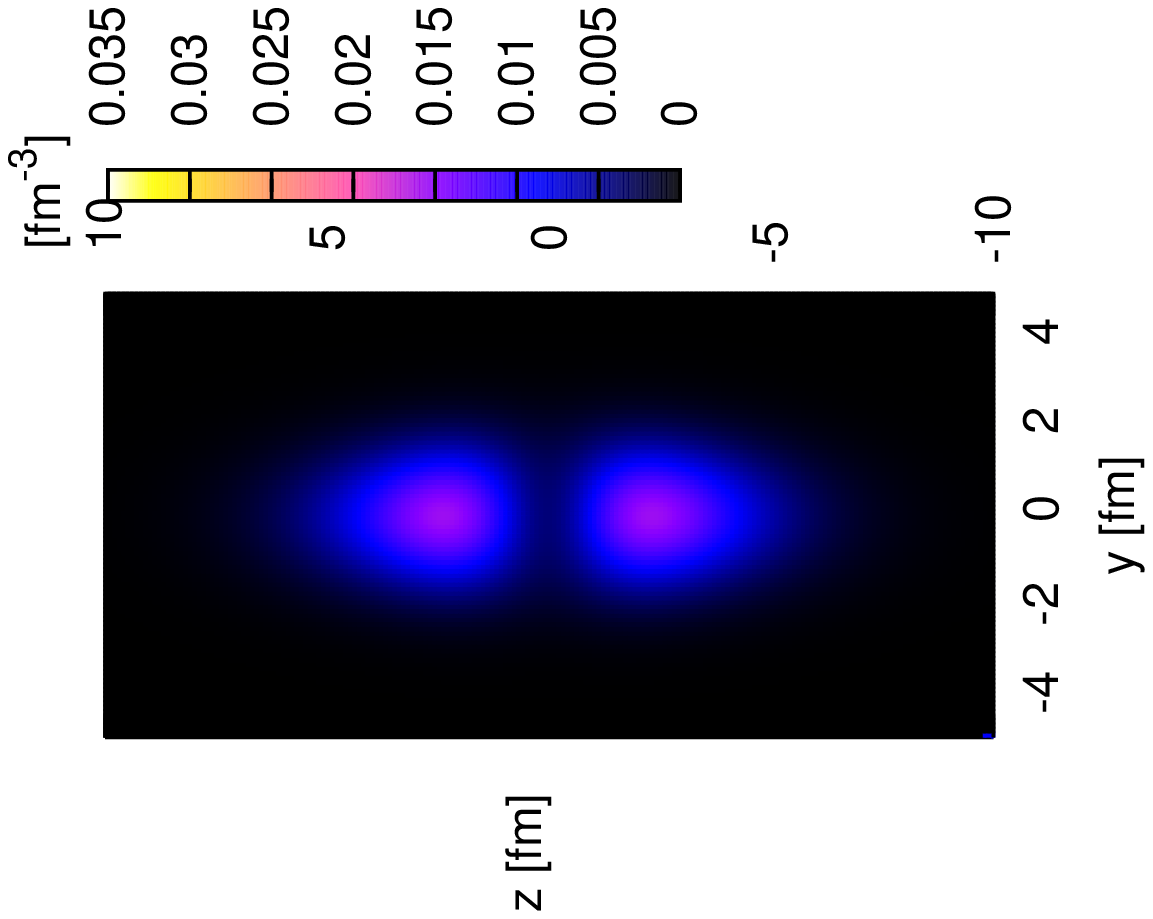}
\caption{ Comparison of single Brink and THSR intrinsic densities for $^8$Be, from top to bottom.}
\label{densities8Be}
\end{figure}

The $\alpha$ clusters can be considered as ideal bosons moving in their own bosonic mean field freely over the whole nuclear volume except for mutual overlaps. This can also nicely be seen in Fig.~\ref{densities8Be} of \cite{Suzuki} where the two $\alpha$ correlation in $^{12}$C($0_2^+)$ as function of their mutual distance is displayed . It practically corresponds to the 'excluded volume' idea often employed phenomenologically in cluster physics. This repulsive 'force' between two $\alpha$'s also is the reason why they cannot be 100 $\%$ in a condensate state but to a certain percentage the $\alpha$'s are scattered out of the condensate. A nice comparison between the Brink and THSR approach is shown in Fig.~\ref{densities8Be} where we compare the intrinsic density distribution of $^8$Be calculated  with the single THSR wave function (lower panel) with the one of a single Brink wave function (upper panel). The strong difference in localization of both distributions should be appreciated. Here the 'intrinsic density' means that the system is in a symmetry broken deformed state which is close to a classical picture. Of course, the ground state (remember that $^8$Be is slightly unstable with a width of only some eV whereas nuclear energy scales are MeV) of $^8$Be has quantum number $0^+$ and in the laboratory frame this state is spherical. This is obtained from the deformed intrinsic state in averaging it over the whole angular range in space.

In more recent works, similar results to the ones with a single Brink wave function have been obtained with a more general mean field approach of the Gogny or Relativistic Mean Field (RMF) type \cite{Girod}. As an example it was found that expanding a nucleus like $^{16}$O employing a constrained Hartree-Fock-Bogoliubov (HFB) approach, at some critical low density, the nucleons spontaneously cluster into a tetrahedron of four $\alpha$ particles. These $\alpha$ particles have fixed positions like they can be formed with a single Brink wave function $\Phi_{\mbox{B}}$ in (\ref{Uegaki}), see Fig. \ref{densities8Be} upper panel. The general mean field approach has, however, the advantage that realistic density functionals can be used. Whether a GCM calculation can be applied on top of these configurations like with a Brink-GCM wave function, remains to be seen.

\subsection{Rotating triangle versus extended THSR approach for the 'Hoyle band'}

\begin{figure}
\includegraphics[width=6cm]{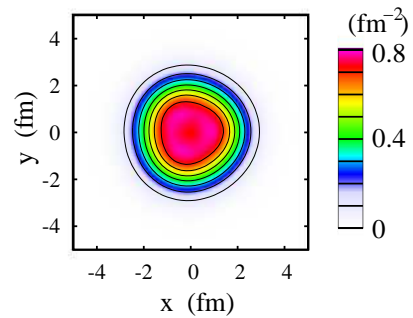}
\caption{\label{Yoshiko}
Intrinsic density distribution of the $^{12}$C ground state from a mean field calculation (we thank Y. Kanada-En'yo for providing this figure).}
\end{figure}

\begin{figure}
\includegraphics[width=7.5cm]{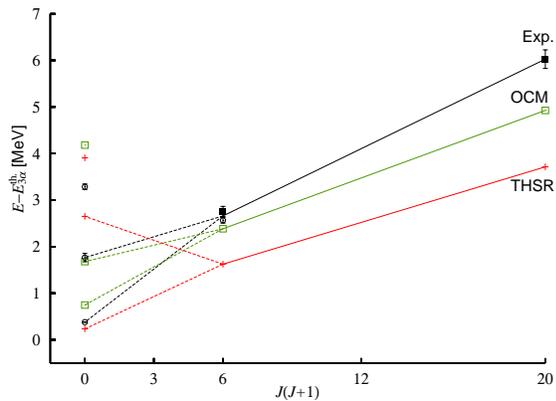}
\caption{\label{hoyle-rot}
Positions of $0^+$ states together with $2^+$ and $4^+$ states of the Hoyle band as a function of $J(J+1)$. The origin at the vertical axis is the 3$\alpha$ disintegration threshold. OCM and extended THSR results are compared with experiment.}
\end{figure}

\noindent
Another approach of $\alpha$ clustering in $^{12}$C has been put forward recently. In \cite{Freer} an algebraic model  put forward by  Iachello {\it et al.} \cite{Iachello}, originally due to Teller \cite{Teller}, was used on the hypothesis that the ground state of $^{12}$C has an equilateral triangle structure. The model then allows to calculate the rotational-vibrational (rot-vib) spectrum of three $\alpha$ particles. Notably a newly measured 5$^-$ state very nicely fits into the rotational band of a spinning triangle. This interpretation is also reinforced by the fact that for such a situation the 4$^+$ and 4$^-$ states should be degenerate what is effectively the case experimentally. In Fig.\ref{Yoshiko}, we show the triangular density distribution of the $^{12}$C ground state obtained from a pure mean field calculation.  This means a calculation without any projection on parity nor angular momentum. Therefore, symmetry is spontaneously broken into a triangular shape. The calculation is obtained under the same conditions as in \cite{Suharo}. However, in that work only figures with variation after projection are shown \cite{Suharo}. This enhances the triangular shape. The Fig.\ref{Yoshiko} is unpublished. It must be said, however, that the broken symmetry to a triangular shape is very subtle and depends on the force used \cite{Yoshiko2}. Anyway, such a triangular shape seems definitely a possibility. The authors in \cite{Freer} then tried to repeat their reasoning tentatively for the 'rotational' band with the Hoyle state as the band head. However, in this case, the situation is much less clear. In Fig.~\ref{hoyle-rot} we see the experimental positions of the $0^+$-states together with the $2_2^+$ and $4_2^+$ ones plotted as a function of $J(J+1)$ and compared with the results of an OCM approach \cite{Ohtsubo} and of  a calculation by Funaki where a generalised THSR wave function has been used \cite{Funaki2} involving a different $B$ parameter for each Jacobi coordinate

\begin{figure}
\includegraphics[width=6cm]{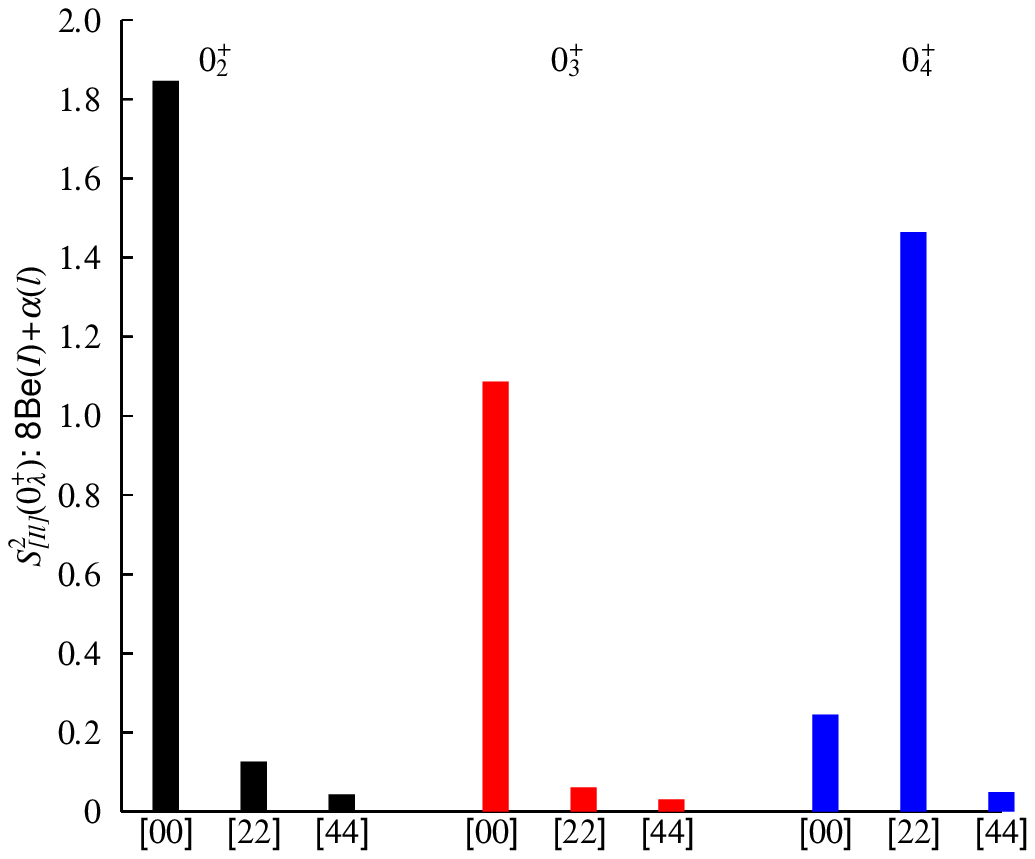}
\includegraphics[width=6cm]{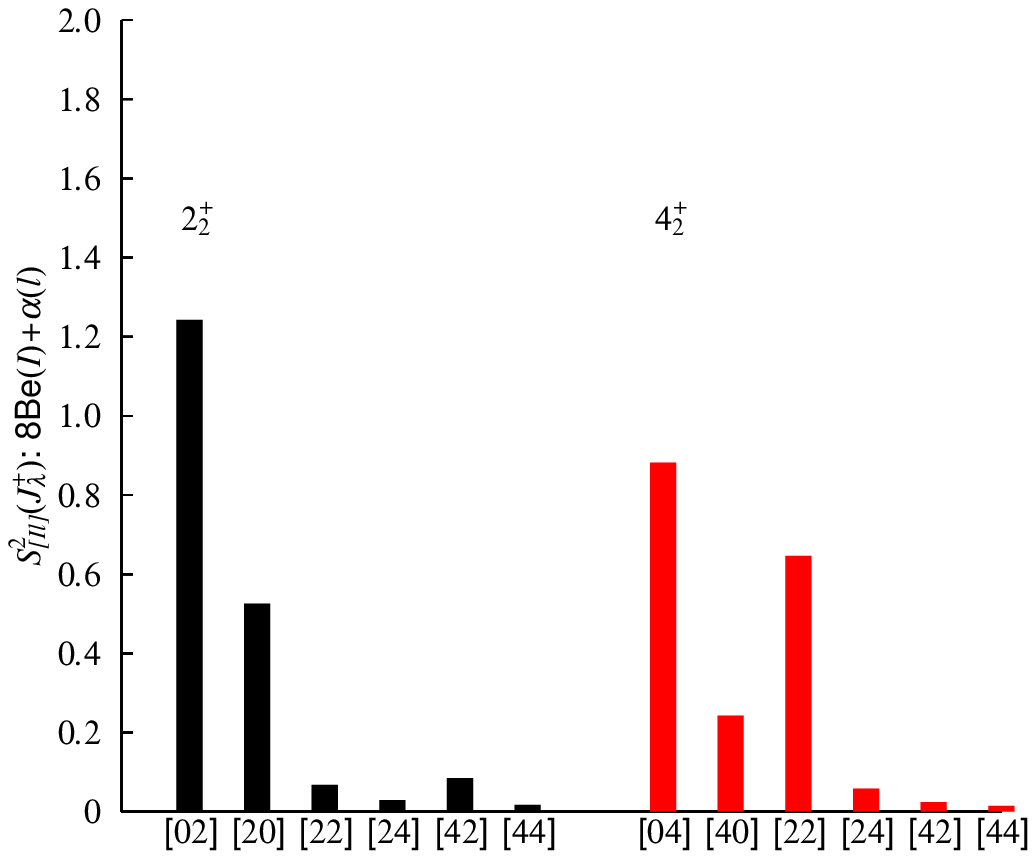}
\caption{\label{probas-0,2,4-hoyle}
Probability distributions for various components in the Hoyle and excitations of the Hoyle state.}
\end{figure}

\begin{equation}
\Psi_{3\alpha} \propto {\mathcal A}\bigg [ \exp \bigg ( -\frac{4}{3B_1^2}{\bf \xi}_1^2 -\frac{1}{B_2^2}{\bf \xi}_2^2 \bigg ) \bigg ] \phi_{\alpha_1}\phi_{\alpha_2}\phi_{\alpha_3}
\label{e-thsr}
\end{equation}

This physically very transparent 12 nucleon wave function, obviously allows to describe pairs of $\alpha$'s to have different relative distances. That is, this generalisation of THSR includes $\alpha$ pair correlations. This is a very important new feature of the THSR approach. With $B_1 =B_2$, one recovers (\ref{THSRwf}).  With (\ref{e-thsr}), a variety of excited $\alpha$ gas states above the Hoyle state have been obtained in \cite{Funaki2}. The result is that the Hoyle state (0$_2^+$) and the third zero plus state (0$^+_3$) have with $B(E2;2_2^+ \rightarrow 0_2^+)= 295$ e$^2$fm$^4$ and $B(E2;2_2^+ \rightarrow 0_3^+)$ = 104 e$^2$fm$^4$, respectively, both a strong transition probability to the second two plus state (2$^+_2$). So no clear band head can be identified. It was also concluded in  Ref. \cite{Richter} that the states $0_2^+, 2_2^+, 4_2^+$ do not form a rotational band. The line which connects the two other hypothetical members of the rotational band, see Fig~\ref{hoyle-rot}, has  a slope which points to somewhere in between of the 0$^+_2$ and 0$^+_3$ states. To conclude from there that this gives raise to  a rotational band, may be premature. One should also realise that the $0_3^+$ state is strongly excited from the Hoyle state by monopole transition whose strength is obtained from the extended THSR calculation to be $M(E0;0_3^+ \rightarrow 0_2^+)$ = 35 fm$^2$. So, the $0_3^+$ state seems to be a state where one $\alpha$ particle has been lifted out of the condensate to the next higher S level with a node, see also Kurokawa {\it et al.} ~\cite{Kurokawa} and Ohtsubo {\it et al.}~\cite{Ohtsubo}, and also ~\cite{YamadaC13}), where the $0^+_3$ and $0^+_4$ states have been identified as well. This is confirmed in Fig.~\ref{probas-0,2,4-hoyle} where the probabilities, $S^2_{[I,l]}$, of the third $\alpha$ orbiting in an $l$ wave around a $^8$Be-like, two $\alpha$ correlated pair with relative angular momentum $I$, are displayed. One sees that except for the $0_4^+$ state, all the states have the largest contribution from the $[0,l]$ channel. So, the picture which arises is as follows: in the Hoyle state, the three $\alpha$'s are all in relative 0S states with some $\alpha$-pair correlations (even with $I\ne 0$, see, e.g., \cite{Rimas, Ishikawa}), responsible for emptying the $\alpha$ condensate by 20-30$\%$. This S-wave dominance, so far found by about half a dozen  different theoretical works, see, e.g., \cite{Horiuchi, Kami, Uegaki, Suzuki, occ's,  Rimas, Ishikawa}, is incompatible with the picture of a rotating triangle. As mentioned, the $0_3^+$ state is one where an $\alpha$ particle is in a higher nodal S state and the $0_ 4^+$ state is built out of an $\alpha$ particle orbiting in a D-wave around a (correlated) two $\alpha$ pair, also in a relative 0D state, see Fig.\ref{probas-0,2,4-hoyle}. The $2_2^+$ and $4_2^+$ states are a mixture of various relative angular momentum states (Fig.~\ref{probas-0,2,4-hoyle}). Whether they can be qualified as members of a rotational band or, may be, rather of a vibrational band or a mixture of both, is an open question.
In any case, indeed,  they are very strongly connected by $B(E2)$ transitions: $B(E2;4_2^+ \rightarrow 2_2^+) = 560$ $e^2$fm$^4$. Let us also mention that the excited $\alpha$ cluster states discussed above have a width much larger ($\sim$ 1 MeV) than the Hoyle state ($\sim$ 1 eV). Nevertheless, they are treated in bound state approximation.

One may also wonder why, with the extended THSR approach, there is a relatively strong difference between the calculated and experimental, so-called Hoyle band?  This may have to do with a deficiency inherent to the THSR wave function which so far has not been cured ( there may be ways to do it in the future). It concerns the fact that with THSR (as, by the way, with the Brink wave function), it is difficult to include the spin-orbit potential. This has as a consequence that the first $2^+$ and first $4^+$ states are quite wrong in energy because the strong energy splitting between $p_{3/2}$ and $p_{1/2}$ states is missing. This probably has a repercussion on the position of the second $2^+$ and $4^+$ states. This can be deduced from the OCM calculation by Ohtsubo {\it et al.} ~\cite{Ohtsubo} also shown in Fig.\ref{hoyle-rot} where the  $2^+$ and $4^+$ states of the ground state rotational band have been adjusted to experiment with a phenomenological force and, thus, the position of the  $2^+$ and $4^+$ states of the so-called Hoyle-band is much improved. Aditionally, this may also come from the fact that with this extended THSR wave function  a different force has to be adopted. Such investigations are under way.

\subsection{Quantum Monte Carlo and {\it ab initio} approaches}

Very recently a break through in the description of the Hoyle state was achieved by two groups \cite{Wiringa, Meissner} using Monte Carlo techniques. In \cite{Meissner} Dean Lee {\it et al.} reproduced the low lying spectrum of $^{12}$C, including the Hoyle state, very accuratly with a so-called ab initio  lattice QMC approach starting from effective chiral field theory. The sign problem has been circumvented exploiting the fact that SU(4) symmetry is very well fullfilled, at least for the lighter nuclei. This parameter free first principle calculation is an important step forward in the explanation of the structure of $^{12}$C. 
On the other hand, all quantities which are more sensitive to details of the wave function have so far either not been calculated (e.g., inelastic form factor to the Hoyle state) or the results are in quite poor agreement with the results of practically all other theoretical approaches. This, for instance, is the case for the rms radius of the Hoyle state which in \cite{Meissner} is barely larger than the one of the ground state whereas it is usually believed that the Hoyle state is quite extended. The authors of \cite{Meissner} remark themselves that higher order contributions to the chiral expansion have to be included to account for the size of the Hoyle state.

\begin{figure}
\includegraphics[width=6cm,angle=-90]{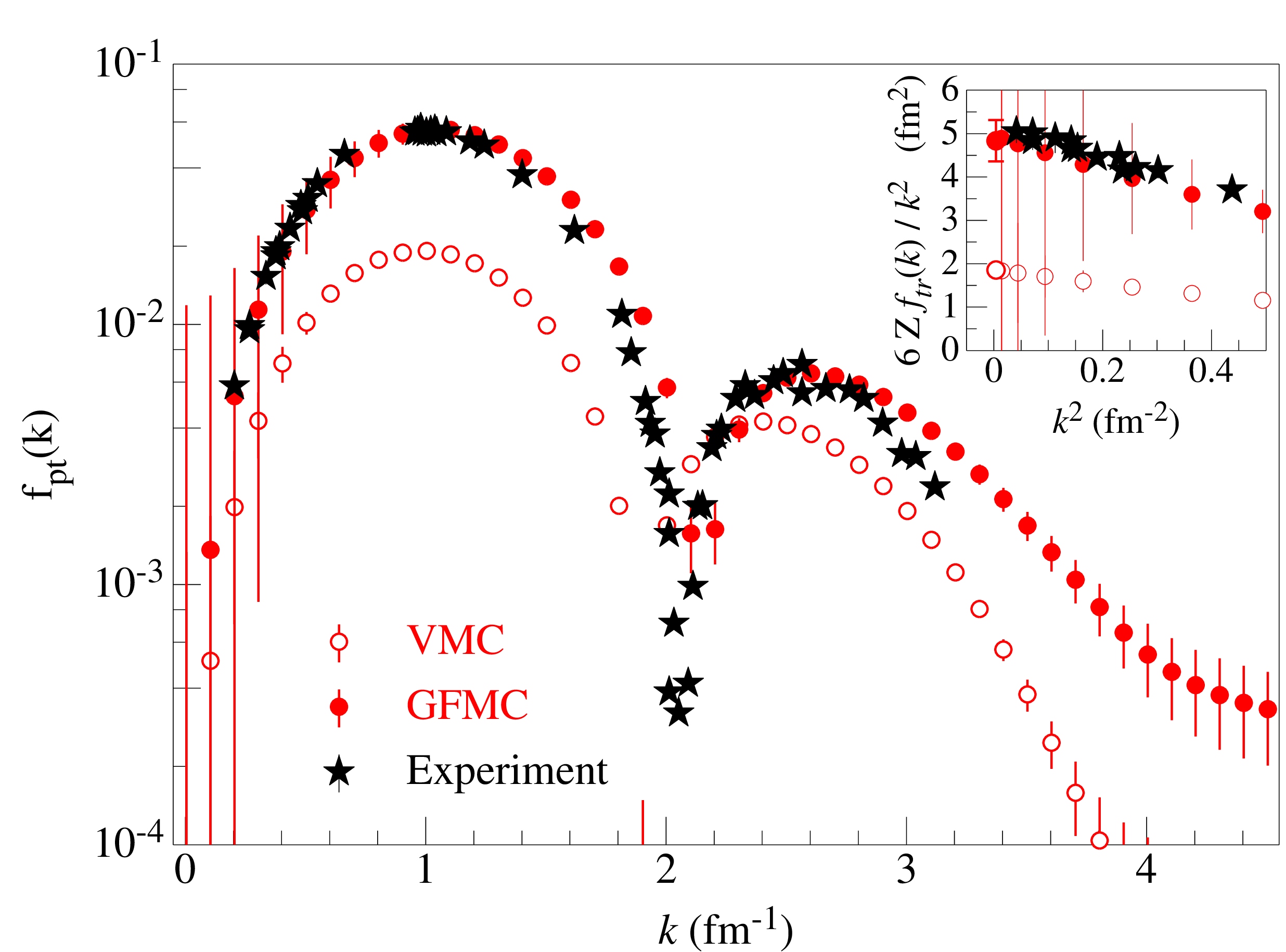}
\includegraphics[width=7cm]{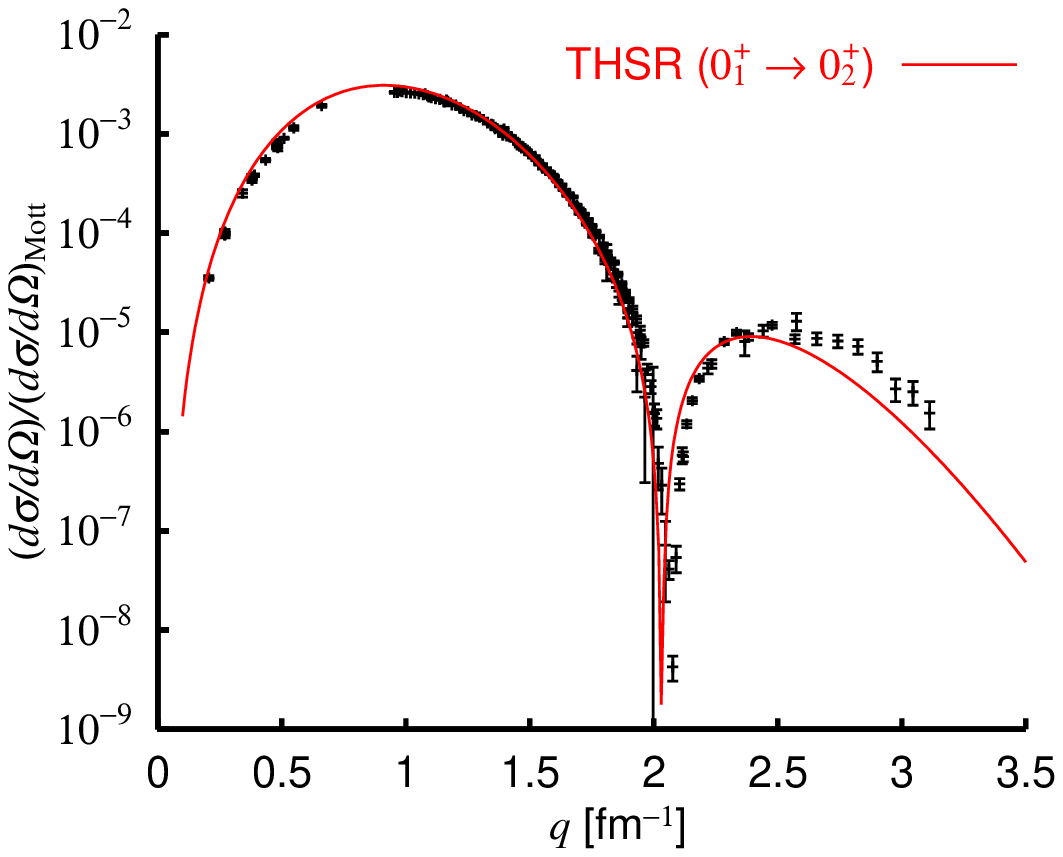}
\caption{\label{QMC}
Inelastic form factors from GFMC ~\cite{Wiringa}, upper panel, and THSR \cite{inelastic}, lower panel. The THSR result cannot be distinguished from the one of \cite{Kami} on the scale of the figure.}
\end{figure}

On the other hand, there exist new Green's function Monte Carlo (GFMC)
results
with constrained path approximation using the Argonne v18 two-body and
Illinois-7 three-body forces, where the inelastic form factor for most of
the
experimental points is reproduced very accurately \cite{Wiringa}. In Fig.~\ref{QMC}, we compare this result with the one obtained from THSR. The results of  Kamimura \cite{Kami} are on top of the THSR ones. They can not be distinguished from the THSR ones on the scale of the graph demonstrating again the equivalence of both approaches. We see excellent agreement between the three calculations and with experiment. In the insert of the upper panel, we see nevertheless that the rather precise experimental transition radius of 5.29$\pm$ 0.14 fm$^2$ given in \cite{Richter} is much better reproduced than in $\alpha$ cluster models which all yield an about 20\% too large value. This may also be the reason for the too slow drop off of the THSR density in the surface region, see lower panel of Fig.\ref{Hoyle-density} below. The energy of the Hoyle state is with around 10 MeV in \cite{Wiringa}  slightly worse than the one in \cite{Meissner}. In Fig.~\ref{Hoyle-density}, we compare the density of the Hoyle state (weighted with $r^2$) obtained with the THSR wave function and in \cite{Wiringa}. We again see quite good agreement between both figures up to about 4 fm. For instance the kind of plateau between 1.5 and 4 fm seems to be very characteristic. It is, however, more pronounced in the GFMC calculation than from THSR. For a better appreciation, we repeat the results of THSR separately in the lower panel of Fig.\ref{Hoyle-density}. Beyond 4 fm, the density in \cite{Wiringa} falls off more rapidly. As already mentioned, this may be due to the fact that the GFMC results are more accurate for small $q$-values. At any rate, the outcome of the three calculations in \cite{thsr, Kami, Wiringa} is so close that it is difficult to believe that results for other quantities should be qualitatively different when calculated with the GFMC technique. This should, for instance, hold for the strong proportion of relative S-waves between the $\alpha$'s found with the other approaches discussed above.

\begin{figure}
\includegraphics[width=6.5cm]{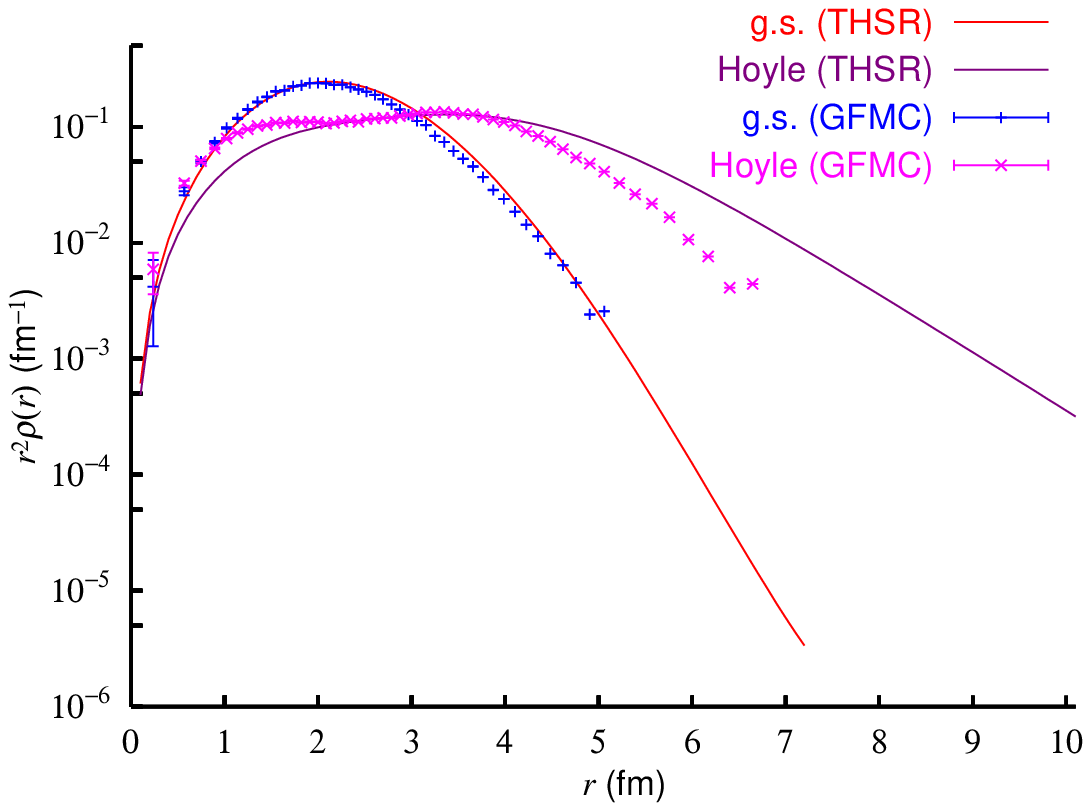}
\includegraphics[width=6.5cm]{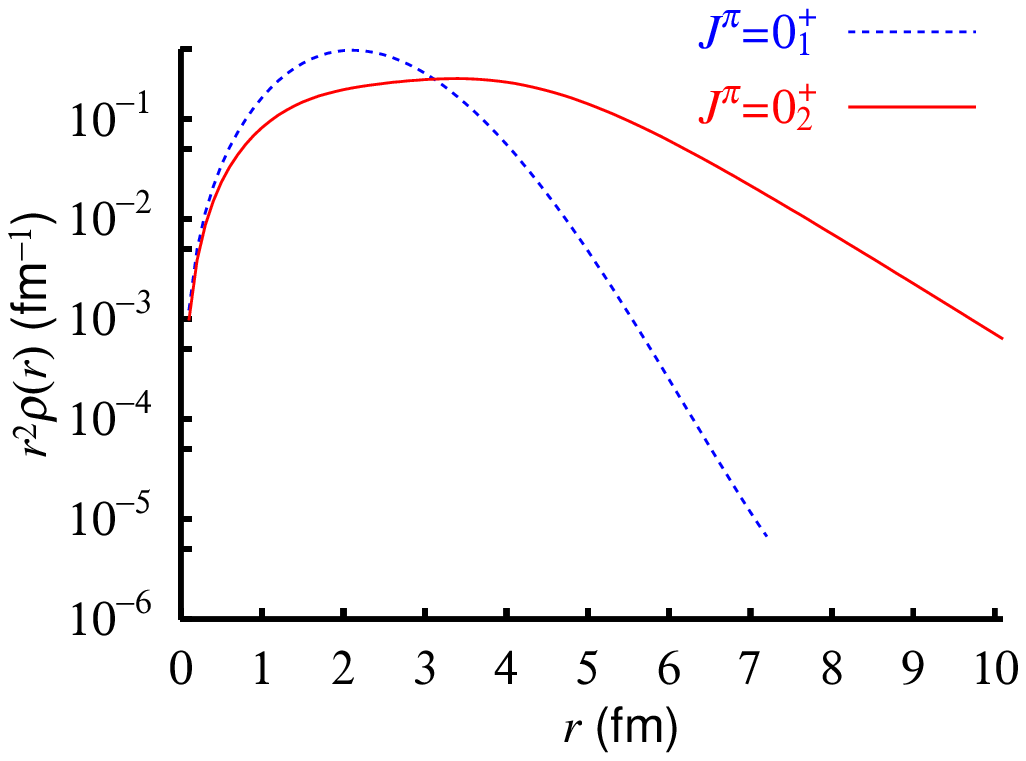}
\caption{Density of the Hoyle state with GFMC ~\cite{Wiringa}, upper panel, and THSR, lower panel. In the upper panel, the diamonds correspond to the ground state density and the full circles to the one of the Hoyle state. The open circles correspond to some approximate calculation, see \cite{Wiringa}.}
\label{Hoyle-density}
\end{figure}

\subsection{Nuclear Matter}

\begin{figure*}\begin{center}
\includegraphics[width=14cm]{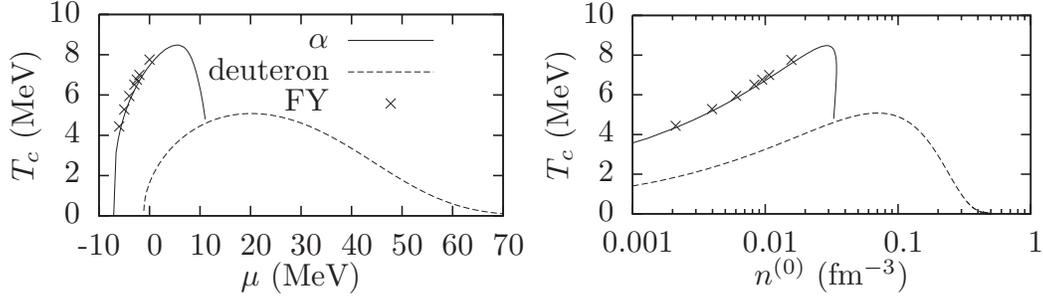}
\end{center}
\caption{Critical temperature for $\alpha$ condensation as a function of chemical potential (left panel) and as a function of uncorrelated density (right panel) compared to the one of neutron-proton (deuteron) pairing (broken line). The crosses correspond to a full solution of the in medium Faddeev-Yakubovsky equations with the Malfliet-Tjohn potential \cite{Sogo} }
\label{Tcrit}
\end{figure*}

Last but certainly not least, we want to consider $\alpha$ clustering and $\alpha$ condensation in nuclear matter. As a matter of fact, it was for nuclear matter where the possibility of $\alpha$ particle condensation had been considered first, see 
\cite{Roepke} where the critical temperature from an in medium four nucleon (two protons and two neutrons) equation has been established. In \cite{Sogo} an improved calculation is presented, see Fig.~\ref{Tcrit}.

 This is in complete analogy to what is known as the Thouless criterion for the onset of pairing as a function of temperature. It was found, that despite of its strong binding, the $\alpha$ condensate, as a function of increasing density, breaks rapidly down as soon as the chemical potential passes substantially from negative values (binding) to positive ones. This is contrary to what happens for pairing where the strong coupling limit passes continuously to the weak, BCS-type of limit with positive values of the chemical potential and a long coherence length (size) of the Cooper pairs ~\cite{Baldo}. The density where $\alpha$ condensation as a function of temperature breaks down is about a fifth of the saturation density. 
 
The reason for this very different behavior between pairing and quartetting has to do with the fact that the in medium two particle level density $g_2(E)= \int \frac{d^3k}{(2\pi\hbar)^3}\delta(E-2\epsilon_k)$, for the two particles at rest, has a finite value at the Fermi level whereas this is not the case with the four body level density
\begin{eqnarray}
g_4(E)&=&\frac{1}{(2\pi \hbar)^{12}}\int d^3k_1d^3k_2d^3k_3d^3k_4 \delta({\bf k}_1 +{\bf k}_2+{\bf k}_3 + {\bf k}_4)\nonumber\\
&\times& \theta_{1234}\delta(E-\epsilon_{k_1}-\epsilon_{k_2}-\epsilon_{k_3}-\epsilon_{k_4})
\end{eqnarray}     
which goes through zero at the Fermi energy, just at the point where quartet correlations should build up. Here $\epsilon_k$ are kinetic energies and $\theta_{1234} =\theta_1\theta_2\theta_3\theta_4 +\bar \theta_1\bar \theta_2\bar \theta_3\bar \theta_4$ and $\theta_i = \theta(\mu -\epsilon_{k_i}), \bar \theta_{k_i} = \theta(\epsilon_{k_i} - \mu)$ with $\mu$ the chemical potential. We leave it to the reader to verify this but it is also explained in \cite{Sogo}. As a matter of fact all many body level densities of this kind go through zero at the Fermi level, besides, precisely in the one body and two body cases when the two particles are at rest. Another well known example of this kind is the two particle-one hole level density which enters the perturbative calculation of the mean free path of a fermion in a fermionic medium. At the Fermi energy the mean free path becomes infinite because the 2p-1h level density goes through zero there.\\ 

In conclusion, it is legitimate to see in the Hoyle state (and other similar states in heavier self-conjugate nuclei) the precursor of the infinite matter situation, see also \cite{Takemoto} where $\alpha$ matter was investigated with a crystal structure. The situation is then quite analogous to pairing in nuclei which can be considered as a precursor of pairing in neutron matter, i.e., neutron stars.

\section{A glimpse on $^{16}$O}

\begin{figure}
\includegraphics[width=6cm]{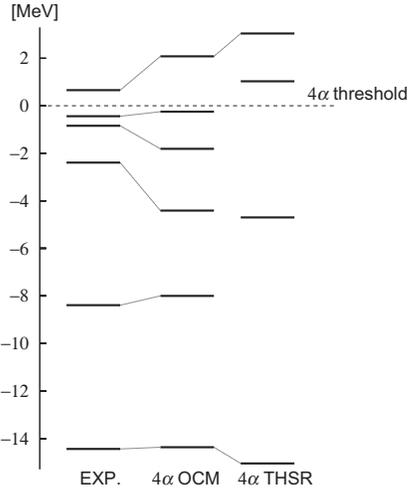}
\caption{\label{16O-spec}
Spectrum of $0^+$ states in $^{16}$O from the OCM approach \cite{OCM16}.}
\end{figure}

The situation in $^{16}$O is again quite a bit more complicated than in $^{12}$C. The fact is that between the 4$\alpha$ threshold and the ground state, there are a couple of $0^+$ states which can be interpreted as $\alpha + ^{12}$C cluster configurations. In Fig.~\ref{16O-spec}, we show the result of an OCM calculation with a very large dimension \cite{OCM16}. 

\begin{figure}
\includegraphics[width=6cm]{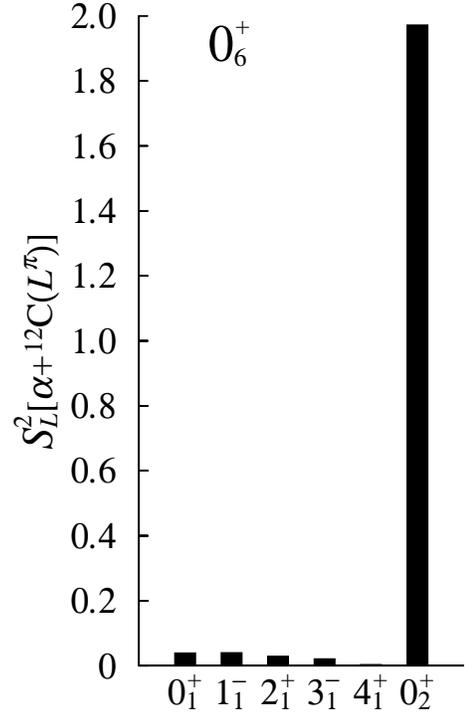}
\caption{\label{spec-factor-16O}
Spectroscopic factor of $0_6^+$ state in $^{16}$O with respect to $\alpha$ plus Hoyle state \cite{yamada-beck}.}
\end{figure}

We see that there is a very nice one to one correspondence between the first six calculated  $0^+$ states and experiment. In regard of the complexity of the situation the agreement between both can be considered as very satisfactory. Only the highest state was identified with the 4$\alpha$ condensate state. The four other excited $0^+$ states are $\alpha +  ^{12}$C configurations. For example the 5-th $0^+$ state is interpreted as an $\alpha$ orbiting in a higher nodal S-wave around the ground state of $^{12}$C. The 4-th $0^+$ state contains an $\alpha$ orbiting in a P-wave around the first $1^-$ state in $^{12}$C. In the 3-rd $0^+$ state the $\alpha$ is in a D-wave coupled to the $2_1^+$ state of $^{12}$C and in the 2-nd $0^+$ state the $\alpha$ is in a 0S-wave and the $^{12}$C in its ground state. The single parameter THSR calculation can only reproduce correctly the ground state and the $\alpha$ condensate state ($0_6^+$). By construction it cannot describe $\alpha + $ $^{12}$C configurations. So, the two intermediate states give some sort of average picture of the four $\alpha$ plus $^{12}$C configurations. One would have to employ a  more general ansatz like in (\ref{e-thsr}) to cope with the situation. Work in this direction is in progress. The $0_6^+$ state is theoretically identified as the $\alpha$-condensate state from the overlap squared $|\langle 0_6^+|\alpha + ^{12}\mbox{C}(0^+_i)\rangle |^2$ \cite{yamada-beck}. In Fig.~\ref{spec-factor-16O} we see that the $0_6^+$ state has an overwhelming contribution from the Hoyle state plus $\alpha$ particle. 

It would be very important to measure, as is the case in $^{12}$C for the Hoyle state, the inelastic form factor from ground to the $0_6^+$ state to have at least an indirect confirmation of a Hoyle-analog state in $^{16}$O.

\section{Experimental evidences ?}

Unfortunately, contrary to pairing, the experimental evidences for $\alpha$ condensation are very  rare and, so far, only indirect. We, nevertheless, want to elaborate here on this issue, even though the experimental situation concerning $\alpha$ particle condensation is far from being clear. However, this may incite experimentors to perform more extensive and more accurate measurements.\\
The most prominent feature is the inelastic form factor which, as stated above, is very sensitive to the extension of the Hoyle state and shows that the Hoyle state has a volume 3-4 times larger than the one of the ground state of $^{12}$C. A state at low density is, of course, very favorable to $\alpha$ condensation as we have seen from the infinite matter study. Nevertheless, this does not establish a direct evidence.  Also the analysis of hadronic  reactions indicate an increased radius of the Hoyle state \cite{Ohkubo, Ito}.
Other attempts to search for signatures of $\alpha$ condensate structures are heavy ion collisions around the Fermi energy where a condensate structure may be formed as intermediate state and correlations between the final $\alpha$ particles may reveal this structure.

\begin{figure*}
\begin{center}
\includegraphics[width=8cm]{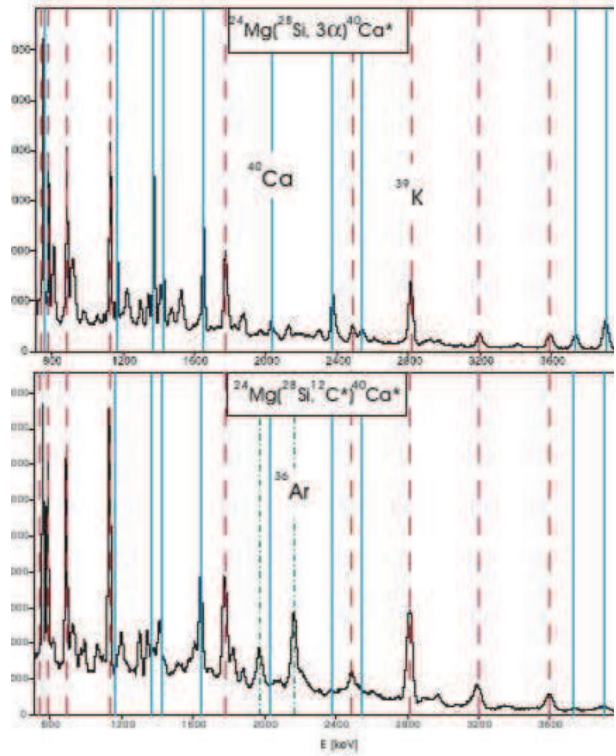}
\caption{ Coincident $\gamma$-spectra gated with the $\alpha$ particles hitting randomly three different detectors (upper panel) in comparison with the case where three $\alpha$'s hit same detector (lower panel). Note the additional lines for $^{36}$Ar in the lower panel.}
\label{36Ar}
\end{center}
\end{figure*}

\begin{figure}
\includegraphics[width=6cm]{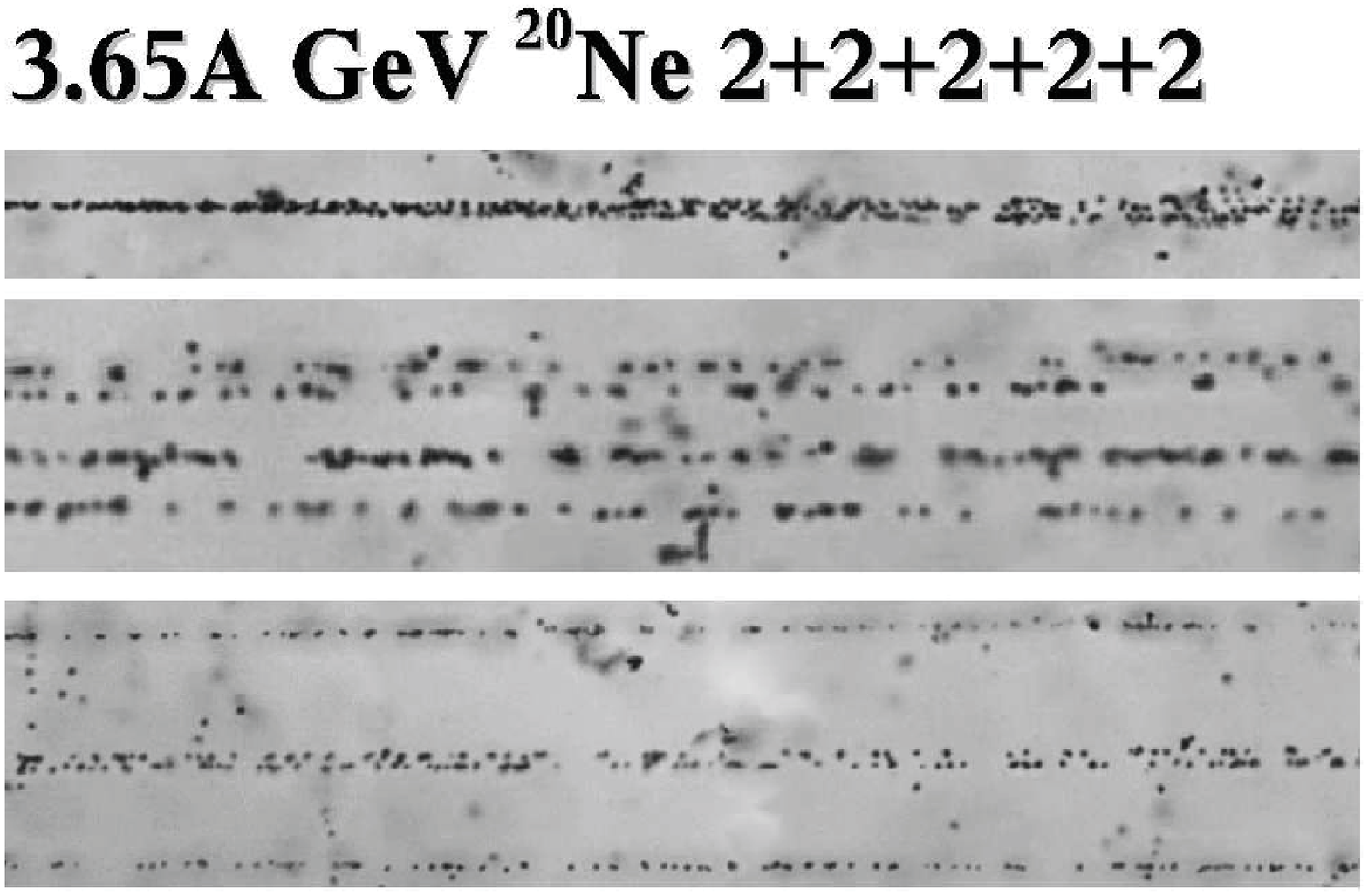}
\caption{Break up of $^{20}$Ne into 5 $\alpha$'s, partially containing a $^8$Be. Figure from \cite{oertzen-beck}.}
\label{Ne20explosion}
\end{figure}

For example von Oertzen {\it et al.} re-analyzed old data ~\cite{Kokalova} of the $^{28}$Si $+ ^{24}$Mg $\rightarrow ^{52}$Fe $\rightarrow ^{40}$Ca $+ 3\alpha$ reaction at 130 MeV which could not be explained with a Hauser-Feshbach approach for the supposedly statistical decay of the compound nucleus $^{52}$Fe. Analyzing the spectrum of the decaying particles via $\gamma$-decay, obtained in combination with a multi-particle detector, it was found that the spectrum is dramatically different for events where the three $\alpha$'s are emitted randomly hitting various detectors under different angles from the ones where the three $\alpha$ were impinging on the same detector. This is shown in Fig.~\ref{36Ar} where the upper panel corresponds to the case of the 3$\alpha$'s in different detectors and lower panel, 3$\alpha$'s in same detector. A spectacular enhancement of the $^{36}$Ar line is seen in the lower panel. This is then explained  by a strong lowering of the emission barrier, due to the presence of an $\alpha$ gas state, for the emission of $^{12}$C($0^+_2$). This fact explains that the energies of the $^{12}$C($0^+_2$) are concentrated at much lower energies as compared to the summed energy of 3$\alpha$ particles under the same kinematical conditions ~\cite{oertzen-beck}. In this way, the residual nucleus ($^{40}$Ca) attains a much higher excitation energy which leads to a subsequent $\alpha$ decay and to a pile up of $^{36}$Ar in the $\gamma$ spectrum. One could also ask the question whether four $\alpha$'s have not been seen in the same detector. However, this only will happen at somewhat higher energies, an important experiment to be done in the future. 

The interpretation of the experiment is, thus, the following, we cite v. Oertzen \cite{oertzen-beck}: {\it 'due to the coherent properties of the threshold states consisting of $\alpha$ particles with a large de-Broglie wave length, the decay of the compound nucleus $^{52}$Fe did not follow the Hauser-Feshbach assumption of the statistical model: a sequential decay and that all decay steps are statistically independent. On the contrary, after emission of the first $\alpha$ particle, the residual $\alpha$ particles in the nucleus contain the phase of the first emission process. The subsequent decays will follow with very short time delays related to the nuclear reaction times. Actually, a simultaneous decay can be considered. Very relevant for this scenario is, as mentioned, the large spacial extension  of the Bose condensate states, as discussed in }~\cite{oertzen-beck}.

\begin{figure}
\includegraphics[width=6cm]{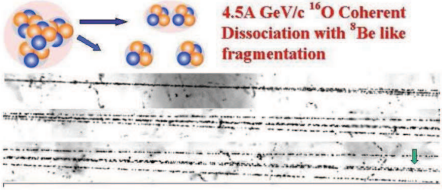}
\caption{Break up of $^{16}$O into 4 $\alpha$'s. Two $\alpha$'s are correlated into $^8$Be. Figure from \cite{oertzen-beck}.}
\label{16Oexplosion}
\end{figure}

However, as the saying goes: 'one swallow does not make a Summer' and, anyway, though suggestive, the above may not be considered as a hard proof of $\alpha$ condensation. It, however, may become a rewarding research field to analyze heavy ion reactions more systematically for non-statistical, coherent $\alpha$ decays.\\
A promising route may also be Coulomb excitation. In Fig.\ref{Ne20explosion}, we show emulsion images of coherent $\alpha$ decay of $^{20}$Ne into three $\alpha$'s and one $^8$Be, or into 5 $\alpha$'s with remarkable intensity from relativistic Coulomb excitation at the Dubna Nucletron accelerator \cite{oertzen-beck}, see also \cite{Bradnova}. The Coulomb break-up being induced by  heavy target nuclei, Silver (Ag).
The break-up of $^{16}$O into 4$\alpha$'s, or into 2$\alpha$'s and one $^8$Be is shown in Fig.\ref{16Oexplosion}. The presence of $^8$Be in the two reactions shows that the $\alpha$'s travel coherently, otherwise the $^8$Be-resonance could not be formed. Of course, also this is only a vague indication for some $\alpha$ particle coherence and much more dedicated experiences should be performed for more firm conclusions.


A dream could be to Coulomb excite $^{40}$Ca to over 60 MeV and observe a slow coherent $\alpha$ particle Coulomb explosion.
Coulomb explosions have been observed in highly charged atomic van der Waals clusters, see ~\cite{explosion}. Coulomb excitation is insofar an ideal excitation mechanism as it transfers very little angular momentum and the projectile essentially gets into a radial density expansion mode.

Next, we want to argue that the $^8$Be decay of the 6th $0^+$ state at 15.1 MeV in $^{16}$O, can eventually show Bose enhancement, if the 15.1 MeV state is an $\alpha$ condensate. 

We know that a pick-up of a Cooper pair out of a superfluid nucleus is enhanced \cite{Broglia}, if the remaining nucleus is also superfluid. As an  example, one could think of the reaction $^{120}$Sn $\rightarrow$ $^{118}$Sn + Cooper pair. Of course same is true for pick up of 2 Cooper pairs simultaneously. We want to make an analogy between this and $^8$Be-decay of 15.1 MeV state. In the decay probability of coincident two $^8$Be, the following spectroscopic factor should enter

\begin{equation}
   S = \langle ^8\mbox{Be} ^8\mbox{Be}|15.1\mbox{MeV}\rangle
\end{equation}
The reduced width amplitude $y$ is roughly related to the spectroscopic factor as  
 $y = 2^{-1/2} (4!/2!2!)^{1/2} S$.
Adopting the condensation approximation of $^8$Be
and 15.1 MeV states, this yields
\[   S=\langle B^2 B^2|(B^+)^4\rangle/(2!2!4!)^{1/2} = (4!/2!2!)^{1/2} = 6^{1/2}\]
entailing $y=6/(2^{1/2}) (y^2=18)$. In above expression for $S$, $B^+ (B)$ stands for an ideal boson creator (destructor), representing the $\alpha$ particle.

When we say that $S$ is large, we need to compare this $S$ with
some standard value.  So we consider the case that the 15.1 MeV state is
a molecular state of $^8$Be-$^8$Be.  We have
\[   S=\langle ^8\mbox{Be(I)} ^8\mbox{Be(II)}|^8\mbox{Be(I)}^8\mbox{Be(II)}\rangle=1 \]
and, therfore, $y=3^{1/2} (y^2=3)$.
This result shows that the condensation character of the 15.1 MeV state
gives a $^8$Be decay width which is 6 times larger than the molecular
resonance character.

We should be aware that above estimate is extremely crude and one rather should rely on a microscopic calculation of the reduced width amplitude $y$ what seems possible to do in the future. Nevertheless, this example shows that the decay of the 15.1 MeV state into two $^8$Be's may be a very rewarding subject, experimentally as well as theoretically, in order to elucidate further its $\alpha$ cluster structure.

A further indirect indication of an extended $\alpha$ gas state and, thus, of the eventual existence of an $\alpha$ condensate state, may be the measurement of the momentum distribution of the $\alpha$ and/or $^8$Be particles from a decaying supposedly $\alpha$ particle condensed state. In ~\cite{occ's} it was shown that those decay products should have a very narrow momentum distribution, close to zero momentum. Again such experiments seem to be very delicate.\\

In conclusion of this section, we may say that the experimental situation needs to be improved. However, new experimental results will soon be published \cite{Natowitz}, or are planned \cite{Freer2}, so that there is hope that we will have a clearer picture of $\alpha$ particle gas states in self-conjugate nuclei in the near future also from the experimental side. 
In this context, we need to mention also two other experimental works. First, there are the results of Raduta {\it et al.} \cite{Raduta}. An enhanced simltaneous 3$\alpha$ decay of the Hoyle state has been found involving a heavy ion reaction. However, this finding is in contradiction with  three other experiments \cite{Itoh, Freer3, Kirsebom} and one theoretical work \cite{Ishikawa} on the decay of an isolated $^{12}$C$^*$ in the Hoyle state where a triple $\alpha$ decay is found to be below the threshold of detectability. It would be important to investigate the reason for this enhanced 3$\alpha$ decay of the Hoyle state in a heavy ion reaction.
Second, there is the recent publication of Marini {\it et al.} \cite{Bonasera} where it is claimed to have detected ``{\it Signals of Bose Einstein condensation and Fermi quenching in the decay of hot nuclear systems}''. In short, in complete vaporisation events, the boson like particles (deuterons, $\alpha$ particles) are much denser packed than the corresponding fermionic particles (protons, $^3$helions, tritons). This would then be in analogy to what has been seen in cold atom systems with fermion-boson mixures \cite{Ebner, Schreck}. We think, however, that much more precise measurements and investigations have to be performed before definite conclusions can be drawn.

Concerning future experiments,
 we would like to repeat that an important quantity still to be measured is the inelastic form factor from ground to the 6th $0^+$ state in $^{16}$O. As we mentioned, this form factor is known since long for the Hoyle state what allowed for strong theoretical conclusions. However, for $^{16}$O this, so far, not possible.

\section{Discussion, Conclusion, and Outlook}

In this short review, we tried to assess the present situation with respect to a possible interpretation of the Hoyle state as an $\alpha$ particle condensate. We pointed to the fact that so far three calculations exist which determine the bosonic occupation numbers of the $\alpha$'s in the Hoyle state \cite{Suzuki, occ's, Ishi-private}. All those works concluded that the three $\alpha$'s of the Hoyle state occupy to $\sim$ 70-80 $\%$ a 0S state with their c.o.m. motion. However, about half a dozen of other works exist which predict a 80 $\%$ relative S-wave dominance between the $\alpha$'s in the Hoyle state, see, e.g., \cite{Horiuchi, Kami, Uegaki, Rimas, Ishikawa}. Since Ishikawa found $\sim 80 \%$ relative S-wave dominance in his 3 boson ($\alpha$) calculation and also calculated the mean field boson ($\alpha$) occupation numbers with also $80 \%$ S-wave, one logically can conclude a strong correlation between dominance of relative S-wave and dominance of S-wave bosonic occupation number. According to this finding, one can say that the Hoyle state is to a large extent an $\alpha$ particle condensate of low density ($1/3-1/4$ of saturation). Quite naturally, this can be considered as a precursor to $\alpha$ particle condensation in low density nuclear matter, see section III.H. This should be seen in analogy to the pairing case where only a handful of Cooper pairs are present and nuclear superfluidity can be considered as the precursor of superfluidity in neutron matter, i.e., neutron stars. The THSR wave function is a single variational wave function which fully respects the Pauli principle among all nucleons and which allows to interpolate between a pure Slater determinant and a pure Bose condensate according to a single variational parameter $B$. We also have considered a hybrid THSR wave function where in the single Brink wave function a variable width parameter $B$ has been introduced. For the positions of the $\alpha$'s all going to zero, one recovers the THSR wave function and for finite positions but $B \rightarrow b$, with $b$ the free space width of the $\alpha$, one recovers the Brink wave function reflecting a crystal structure of the $\alpha$ arrangement. The two variational parameters $B$ and positions have been put into competition with a variational calculation for the energy. The variation largely yields an answer close to the Bose condensate picture, i.e., a large $B$ value, covering the whole nuclear volume, and with positions of the $\alpha$'s all centered at the origin. Such competition has also been analyzed schematically by Zinner and Jensen ~\cite{Zinner} who also concluded that a large extension of the $\alpha$ wave functions covering the whole nuclear volume is akin to Bose condensation. The parameter free reproduction of all experimentally known properties of the Hoyle state with THSR, gives a further strong argument for the condensate picture. Since the THSR wave function has a $\sim$ 98 percent squared overlap with Kamimura's Hoyle state, the former is not just an approximation to the latter but is equivalent. One can, thus,  argue that implicitly the work of Kamimura (and Uegaki) also describes the Hoyle state as a Bose condensate of $\alpha$ particles, a new insight to the otherwise very successful approaches of those authors about 40 years ago. We also pointed out that their work can still to day be considered as the most advanced approach to the $\alpha$ cluster structure of $^{12}$C. Their wave function does not contain any preconceived ingredients for $\alpha$ particle condensation because, in principle, with RGM or Brink-GCM the $\alpha$'s can take any arrangement they like. We surmise that all approaches which so far reproduced the measured properties of the Hoyle state, for instance the inelastic form factor, implicitly describe the same $\alpha$ particle condensate as does the THSR approach. This should notably be the case for AMD and FMD theories and, in particular, also with the very recent GFMC approach. It is nevertheless very desirable that the bosonic occupation numbers will be calculated with those approaches as well.

\begin{figure}
\includegraphics[width=7.5cm]{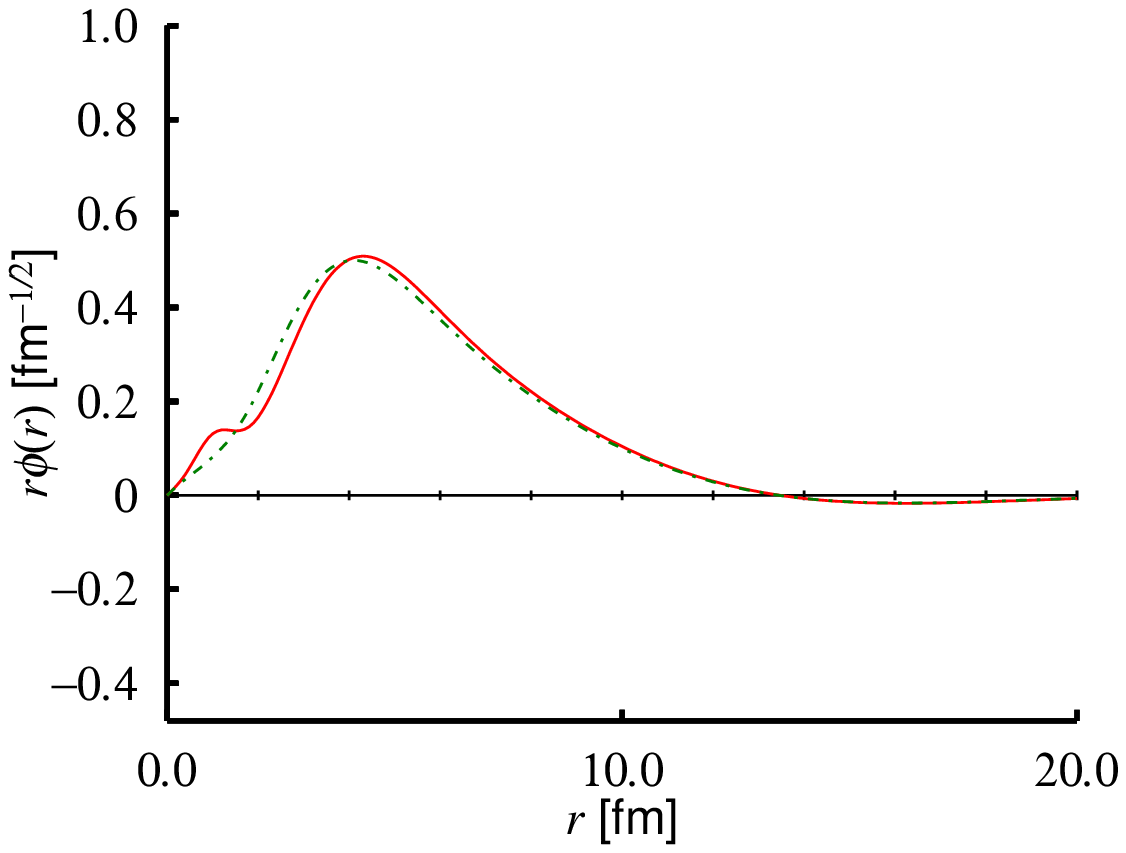}
\includegraphics[width=7.5cm]{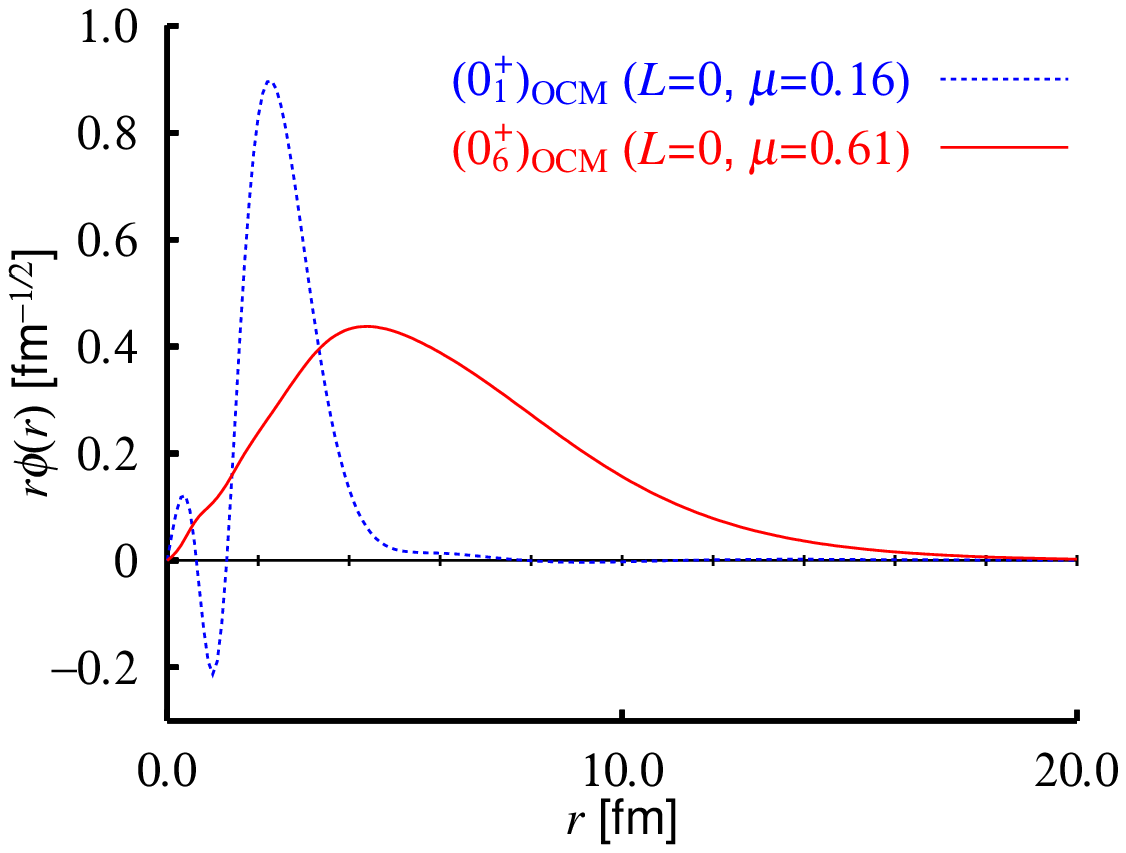}
\caption{\label{alpha-wavefcts} Comparison of single $\alpha$ particle wave functions in the condensate states of $^{12}$C and $^{16}$O. One should remark the similarity of both wave functions (up to a scale factor). The dotted line in the upper panel is a best fit of a Gaussian to the calculated curve (full line). In the lower panel, the dotted line represents the single $\alpha$ partical wave function in the ground state. The strong distortion of an $\alpha$ particle in the compact ground state should be remarked. The upper panel shows the same quantity as in Fig.\ref{gs-wf12C} but from a different calculation  \cite{GPE,yamada-beck}. We show it here again for a direct comparison with the $^{16}$O case.}
\end{figure}

Further indications of the validity of the condensation picture, also discussed in ~\cite{Zinner} are the fact that the de-Broglie wave length of the $\alpha$ particles in the Hoyle state is by factors larger than the extension of the Hoyle state, that is, larger than the inter $\alpha$ distance, Ref. \cite{oertzen-beck}, \cite{GPE}, \cite{yamada-beck}. Also, the calculated shape of the $\alpha$ particle wave function in the condensate practically does not change, besides a trivial norm factor, from $^{12}$C to $^{16}$O, see Refs.~ \cite{GPE,yamada-beck}, this 
being another criterium of $\alpha$ condensation, see Fig.\ref{alpha-wavefcts}.

In this work, we only shortly discussed the situation in $^{16}$O where the $0_6^+$ state at 15.1 MeV is identified as an $\alpha$ condensate state ~\cite{OCM16}. Since the $\alpha$ disintegration threshold rises rather sharply with the number of $\alpha$ particles, one may wonder whether states at such high energies do not acquire a very large width, i.e., decay in very short times. The 15.1 MeV state in $^{16}$O has a width of only 160 keV what is very small considering that excitation energy. This stems from the fact that all the 
states underneath have a strongly different structure. Nevertheless, the ground states have a certain percentage of $\alpha$ gas components and vice versa the condensate states have some shell model components. This gives raise to the decay probability which, of course, increases with more $\alpha$ particles but shall stay unusually small.

Very promising approaches to the Hoyle state are two recent attempts using QMC techniques. Epelbaum {\it et al.} \cite{Meissner} used so-called lattice QMC based on chiral perturbation theory with EFT (Energy Functional Theory). The only open input parameters are the current quark masses. The low lying spectrum of $^{12}$C is very well reproduced. However, no inelastic form factor is calculated as yet. Pieper {\it et al.} \cite{Wiringa} make use of GFMC with the fixed node approximation. The inelastic form factor of the Hoyle state is very well reproduced, see Fig.~\ref{QMC} where we also show the inelastic form factor obtained with both the THSR approach and the one of \cite{Kami}. All three theories reproduce the inelastic form factor very well. 
If at all possible to evaluate, it would be very interesting to see what the GFMC approach yields for the bosonic occupation numbers. On the experimental side, the Bose condensate character is difficult to verify. However, we discussed heavy ion reactions and $^8$Be $+ ^8$Be decay out of the $0^+_6$ state at 15.1 MeV in $^{16}$O as possible future indicators of $\alpha$ particle condensation. Also unusually low momenta of the decay products may give a hint.

All in all, there exist many calculations, see, e.g., \cite{Horiuchi, Kami, Uegaki, Suzuki, occ's, Rimas, Ishikawa, Ishi-private} which all point to the Hoyle state as being dominated by S-waves among the 3$\alpha$'s. We see no counter argument which would invalidate the hypothesis that the Hoyle state is to a large extent composed of an $\alpha$ particle condensate with 70-80 $\%$ occupancy. These results are obtained from sophisticated but natural and transparent wave functions through a Raleigh-Ritz variational principle and the conclusions drawn from these investigations seem to us very reliable. Additionally, there are clear theoretical indications that the 6-th $0^+$ at 15.1 MeV in $^{16}$O is a Hoyle analog state.

\begin{figure}
\includegraphics[width=4.0cm]{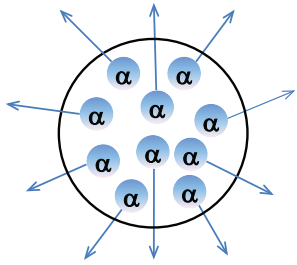}
\caption{Artist's view of a Coulomb explosion of $^{40}$Ca into 10 $\alpha$'s.}
\label{Ca40explosion}
\end{figure}

In this review, we concentrated on the case of the Hoyle state with only a small glimpse on the situation in $^{16}$O. However, it seems clear that in heavier self-conjugate nuclei, like $^{20}$Ne, $^{24}$Mg, up to $^{40}$Ca close to the $\alpha$ disintegration threshold, analogous Hoyle-like $\alpha$ condensates may exist and that a whole series of excited states of which the Bose condensate can be considered as the ground state (see citation of Uegaki above) still is to be discovered and their precise nature to be clarified in the future. Studies in this direction have been performed in \cite{GPE} using the Gross-Pitaevskii equation for bosons. It seems that around $^{40}$Ca the Coulomb barrier fades away and no long lived $\alpha$ condensate can exist any more. The $\alpha$-like correlations and $\alpha$ formation is also of importance for nuclei with $\alpha$ decay like $^{212}$Po \cite{212Po} and superheavy nuclei.  Even the decay of heavier clusters has been observed like $^{223}$Ra into $^{209}$Pb $+ ^{14}$C and discussed theoretically \cite{Bertsch}. $^{20}$Ne is similar to $^{212}$Po with an $\alpha$ particle sitting on top of a doubly magic nucleus ($^{16}$O). In this respect, it is worth pointing out that already mean field approaches (the independent particle model) can show sizeable $\alpha$ cluster correlations \cite{Ebran, Girod}. We have argued that heavy ion reactions with detection of coherent $\alpha$ particle motion have been seen in one or two works in the past \cite{Kokalova} and references in there. However, these reactions seem to be a largely unexploited territory  concerning $\alpha$ particle coherence and condensation. We also pointed out that Coulomb excitation could be an ideal way of inducing important radial extension
of a nucleus provoking ($\alpha$) clustering. A cartoon of a hypothetical Coulomb explosion of $^{40}$Ca into ten $\alpha$'s is shown in Fig.\ref{Ca40explosion}. It seems a truly exciting aspect that in the lighter n$\alpha$ nuclei there is a coexistence of two almost ideal quantum gases: fermions (nucleons) and bosons ($\alpha$ particles). Still many things have to be discovered in this context in future research where nuclear physics plays a prominent role. On the other hand, cluster physics is also very developed concerning atomic clusters \cite{Guet}. However, so far, no bosonic condensation phenomena are discussed in this field, to the best of our knowledge.

\section{Acknowledgements}
We thank very much our collaborators Y. Funaki and T. Yamada who joined the $\alpha$ particle cluster collaboration later but whose contributions have in several aspects been decisive. Also very recent collaborations with Z. Ren, Chang Xu, Mengjiao Lyu, and Bo Zhou are strongly appreciated. P.S. is greatful to P. Nozi\`eres for his continuous interest in quartet condensation. We are greatful to Y. Kanada-En'yo for discussions and for preparing Fig.5 for this article.

\end{document}